\documentclass[manuscript,screen]{acmart}
\AtBeginDocument{%
  }

\usepackage{longtable}   
\usepackage{booktabs}    
\usepackage{array}       
\usepackage{tabularx}  
\usepackage{ragged2e}  
\usepackage[table]{xcolor}
\usepackage{subcaption}
\usepackage[table]{xcolor}

\newif\ifshowrevisions
\showrevisionsfalse
\newcommand{\revise}[1]{\ifshowrevisions\textcolor{red}{#1}\else#1\fi}

\setcopyright{acmlicensed}
\copyrightyear{2018}
\acmYear{2018}
\acmDOI{XXXXXXX.XXXXXXX}
\acmConference[Conference acronym 'XX]{Make sure to enter the correct
  conference title from your rights confirmation email}{June 03--05,
  2018}{Woodstock, NY}
\acmISBN{978-1-4503-XXXX-X/2018/06}

\begin{document}

\title{Between Knowledge and Care: A Mixed-Methods Evaluation of Generative AI for T2DM Self-Management from Patient and Physician Perspectives}

\author{Ruiqi Chen}
\authornote{Ruiqi Chen and Yibo Meng contributed equally to this research.}
\email{ruiqich@umich.edu}
\affiliation{%
  \institution{University of Michigan}
  \city{Ann Arbor}
  \state{Michigan}
  \country{USA}
}

\author{Yibo Meng}
\authornotemark[1]
\email{yim4007@med.cornell.edu}
\affiliation{%
  \institution{Weill Cornell Medicine, Cornell University}
  \city{New York}
  \state{New York}
  \country{USA}
}

\author{Huidi Lu}
\email{huidi.lu@sbs.ox.ac.uk}
\affiliation{%
  \institution{University of Oxford}
  \city{Oxford}
  \state{England}
  \country{United Kingdom}
}

\author{Xiaolan Ding}
\affiliation{%
  \institution{North China University of Science and Technology Health Science Center}
  \city{Tangshan}
  \state{Hebei}
  \country{China}
}

\renewcommand{\shortauthors}{Chen et al.}

\begin{abstract}
Generative AI is increasingly used for everyday health guidance, yet its clinical appropriateness in chronic disease contexts remains poorly understood. This paper presents a two-part mixed-methods study on \revise{Type 2 Diabetes Mellitus (T2DM)}, examining how patients and physicians assess AI-generated health information. \revise{Study~1} analyzes 784 \revise{participant reported} patient queries to characterize seven informational need categories and \revise{develops a structured five dimensional physician rating rubric informed by patient query categories and clinician priorities} (\textit{Accuracy, Safety, Clarity, Integrity, Action Orientation}). \revise{Study~2} engages seven physicians scoring responses from four AI models and discussing evaluative reasoning through in-depth interviews. Models perform well on factual explanation and lifestyle guidance but consistently underperform on medication reasoning and emotional support. Two \revise{analytic concepts} emerge \revise{from the data}. The \textit{pre-visit primer} \revise{frames AI as preparation for clinical encounters rather than as a replacement for physicians}. The \textit{fluency illusion} \revise{describes how polished language may convey epistemic authority that the clinical content does not support}. Patients and physicians converged on three shared limitations (role boundaries, emotional inadequacy, personalization gaps) while diverging in evaluative emphasis, \revise{which informed} four design directions, task-aware orchestration, risk-aware fallback, dynamic personalization, and emotionally attuned interaction.
\end{abstract}



\received{20 February 2007}
\received[revised]{12 March 2009}
\received[accepted]{5 June 2009}

\maketitle

\section{INTRODUCTION}
As generative AI systems increasingly permeate consumer health contexts, their potential role in chronic disease management has garnered substantial interest from the HCI and medical informatics communities \cite{cao2022multichannel}. Large language models (LLMs) have demonstrated strong capabilities in synthesizing biomedical knowledge \cite{samimi2025visual,hernandez2023future,ma2026can,zhang2025openhoiopenworldhandobjectinteraction}. Yet, important questions remain about how such systems perform in real-world care settings---particularly with respect to contextual safety, practical actionability, and emotional appropriateness \cite{akrimi2025chatgpt,meng2026balancing,cao2026beyond}.

In this work, we focus on Type 2 Diabetes Mellitus (T2DM) as a representative chronic condition to examine how generative AI tools intersect with everyday patient needs and clinical expectations. T2DM affects over 500 million people globally and requires continuous self-management involving medication, diet, physical activity, and psychosocial regulation \cite{gong2020my}. In contexts such as China---where patient volumes are high and access to endocrinologists is uneven---patients increasingly turn to generative AI tools as always-available sources of health information \cite{hernandez2023future,lanshan2025factors,guo2025promoting}. While our study is situated in this context, the pressures it exemplifies---growing patient reliance on AI for health information alongside constrained specialist consultation time---are visible across a wide range of healthcare systems, making the tensions we document relevant beyond any single setting.

Despite their growing adoption, there remains limited understanding of how current generative AI systems align with the lived realities and risk sensitivities of chronic care. Prior work has predominantly emphasized benchmark-style evaluations or guideline-based validation pipelines \cite{hernandez2023future,akrimi2025chatgpt,antonie2025role,rajagopal2025generativeaisupportpatients}, often abstracted from the situated contexts in which patients interpret and act upon AI-generated advice. More broadly, within HCI, relatively few studies have examined how AI-mediated health information is experienced across the emotional, behavioral, and safety-critical dimensions of long-term disease management \cite{10.1145/3544548.3581251,mirbabaie2025digital,mitchell2021automated,chen2026not}.

To explore these gaps, we conducted a two-part mixed-methods study foregrounding both patient experiences and clinician perspectives. Study~1 analyzes \revise{participant reported prior AI use} among 21 T2DM patients and \revise{784 participant reported patient queries} to characterize seven recurring categories of informational need. Working with physician participants, we then \revise{developed a structured five dimensional physician rating rubric informed by these categories and by clinician priorities}, \textit{Accuracy, Safety, Clarity, Integrity}, and \textit{Action Orientation}. Study~2 engages seven endocrinologists who scored AI-generated responses from four mainstream AI systems using this \revise{rubric} and discussed, through in-depth interviews, the clinical reasoning behind their evaluations. We then analyze where patient and physician perspectives converge and where they diverge \revise{as triangulation across independently collected patient accounts and physician evaluations}.

We operationalize the five dimensions as a structured \revise{rubric} with explicit scoring criteria and differential weighting by clinical priority, while recognizing that its formulation is grounded in the T2DM chronic care context from which it was developed rather than as a universal standard. Findings reveal consistent AI strengths in factual and lifestyle domains alongside significant weaknesses in medication interpretation, contextual reasoning, and emotional support---a profile that holds across both patient-reported experience and physician expert assessment.

Our contributions are threefold. First, we provide \revise{a participant grounded characterization} of T2DM patient informational needs derived from 784 \revise{participant reported} patient queries, establishing seven recurring categories and identifying the \textit{pre-visit primer} as \revise{a GenAI specific form of pre consultation preparation in chronic disease self-management}. Second, we present a \revise{physician informed} five-dimensional evaluation \revise{rubric} and \revise{show that it differentiated physician ratings across models, dimensions, and question categories in this dataset}. Third, we analyze where patient and physician perspectives on AI-generated health information converge and where they diverge, identifying three \revise{shared concerns}, role boundaries, emotional support gaps, and personalization needs, and two divergences in evaluative concern and trust calibration. We also use the term \textit{fluency illusion} to name \revise{a recurring pattern in our data}, the tendency for linguistically polished AI outputs to convey an impression of epistemic authority that the underlying clinical content does not warrant. Together, these findings \revise{inform} four design directions grounded in the performance gaps observed in this study.

We conclude by reflecting on how chronic illness contexts impose particular demands on AI health tools---ones that must move beyond generic question answering toward responses that acknowledge uncertainty, respect clinical boundaries, and adapt to the situated needs of long-term care. By foregrounding both patient and physician perspectives, this work contributes empirically grounded design insights to the HCI community's ongoing discussions on how generative AI might be more safely and thoughtfully deployed in healthcare.

\section{RELATED WORK}

\subsection{Information Needs in T2DM Self-Management}

\revise{T2DM} is an information-intensive chronic condition, requiring patients to make ongoing decisions related to medication, diet, physical activity, and emotional regulation \cite{gong2020my, gerstenberg2025living, cui2025empowering,meng2026tibetcpr}. Prior work in HCI and health informatics has documented persistent challenges patients face in interpreting clinical guidelines, contextualizing numerical indicators such as HbA1c \revise{(glycated hemoglobin, a marker of average blood glucose over roughly the previous two to three months)}, and translating abstract medical knowledge into everyday self-management \cite{10.1145/3581641.3584075, hernandez2023future, mayberry2019mhealth, preum2019information}.

In practice, patients frequently supplement clinical consultations with online sources and AI-powered tools, particularly in healthcare systems where consultation time is limited \cite{10.1145/3290605.3300600, peimani2024moderating, hughes2024development}. This pattern is especially pronounced in the period \textit{before} clinical visits, when patients seek to interpret symptoms, prepare questions, and orient themselves within a clinical encounter they have limited time to navigate. Importantly, existing studies show that these information needs extend beyond factual explanation to include lifestyle adaptation and emotional coping \cite{akrimi2025chatgpt, dsouza2024identification, biernatzki2018information}---a breadth that generic information systems have consistently struggled to address.

Within HCI, T2DM has served as a recurring testbed for intelligent health technologies, including education tools, self-tracking systems, and coaching interfaces \cite{bhattacharya2023directive, ayobi2021co, ayobi2023computational,chen2025gestobrush}. These systems demonstrate the value of tailored feedback and interactive support, but typically rely on rule-based logic or pre-scripted content. The emergence of generative AI introduces new conversational flexibility---but also raises open questions about how such systems respond to the safety-sensitive, contextually situated, and emotionally complex concerns that characterize chronic disease self-management.

\subsection{Evaluating AI-Generated Health Information}

Assessing whether LLM-generated content meets the standards required for health contexts has become an active area of concern \cite{10.1145/3613904.3641998, lanshan2025factors, rebitschek2025evaluating, reddy2024generative}. A growing body of evidence indicates that such content may be clinically inaccurate, incomplete, or misleading when applied to domain-specific scenarios such as diabetes care \cite{guo2025promoting, agarwal2024medhalu, vishwanath2024faithfulness, kim2025medical}.

Most existing evaluations rely on benchmark medical question-answering datasets---such as MedQA or PubMedQA---that prioritize factual correctness at the sentence or answer level \cite{antonie2025role, jin2019pubmedqa, jin2021disease, kim2024medexqa}. While valuable for measuring factual precision, these approaches provide limited insight into how patients interpret and act upon AI-generated advice in real-world settings. Within HCI and health informatics, researchers have argued for multidimensional evaluation approaches that account for safety, clarity, empathy, and actionability \cite{10.1145/3544548.3581251, 10.1145/3706598.3713819, kim2024human, rong2023towards}---recognizing that information quality in health contexts cannot be reduced to accuracy alone.

Despite this shift, chronic disease settings remain underexplored. Existing studies in T2DM largely focus on usability or patient satisfaction \cite{akrimi2025chatgpt, swallow2025digibete, tseng2025designing, meng2025between}, with limited involvement of clinicians in evaluating AI-generated content. This matters for design: patient-reported satisfaction and clinician-assessed appropriateness reflect different evaluative standards---one grounded in the experience of the interaction, the other in the clinical consequences of the information. A response that satisfies a patient's immediate informational need may simultaneously fail the clinical safety standards that a physician would apply. Understanding where these evaluative standards converge and where they diverge is essential for designing AI health tools that are simultaneously usable by patients and credible to the clinicians who bear accountability for care.

\subsection{Human--AI Interaction in Chronic Care: Toward Dual-Perspective Evaluation}

Chronic disease management poses distinct challenges for the design of AI-mediated health tools. Unlike acute care scenarios, conditions such as T2DM require sustained informational support over years, involving ongoing decisions about medication adherence, dietary adjustment, and psychosocial regulation \cite{mirbabaie2025digital, mitchell2021automated, mack2022chronically,meng2026living}. These longitudinal demands require not only factual accuracy but also behavioral relevance, emotional sensitivity, and safety awareness that must adapt across a patient's changing health trajectory \cite{oh2023designing, hwang2025effectiveness,meng2026creating,meng2026decoration}.

Prior HCI research has explored a range of patient-facing tools in chronic care---including mHealth apps, digital coaching systems, and shared decision-making interfaces---documenting how patients seek information, build trust in digital sources, and navigate health advice in everyday contexts \cite{oh2023designing, hwang2025effectiveness, chen2023design,su2025flymethrough}. While these systems provide valuable interaction design grounding, they largely predate the widespread adoption of generative AI and thus do not address the specific dynamics introduced by LLM-generated health content.

The emergence of LLMs has intensified both the potential and the risks of AI-mediated health communication. Existing research has proposed risk-sensitive design strategies for AI health tools---such as deference to clinicians, uncertainty disclosure, and adaptive language modulation \cite{schombs2025designing, kandel2025graphical,zhang2026pervasive,meng2026balancing}---but these strategies have largely been developed through conceptual analysis rather than being grounded in clinicians' direct evaluation of real AI-generated outputs across the range of patient concerns that actually arise in chronic disease contexts. Without this clinical grounding, it remains unclear which types of patient questions or response qualities genuinely warrant different system behaviors.

A further gap concerns the relationship between patient and clinician perspectives. HCI research on AI health tools has predominantly centered one group: either patients---studying usability, perceived utility, and trust---or clinicians---assessing accuracy, workflow fit, and clinical appropriateness. Studies that engage both groups simultaneously around the same AI-generated content are rare, and as a result we lack evidence on whether the limitations that patients perceive in AI-generated health information are independently visible to clinicians, or whether each group identifies structurally different gaps. This distinction matters for design: if patients and clinicians converge on the same limitations, those represent high-priority targets for improvement; if they diverge, the design challenge is to serve two different evaluative frameworks simultaneously---not to optimize for one at the cost of the other.

Taken together, these three bodies of work point to a shared gap. Research on T2DM information needs establishes that patients require contextually relevant, emotionally attuned, and behaviorally actionable support---needs that generative AI has not yet been systematically tested against. Research on AI evaluation establishes that multidimensional, clinician-involved assessment is necessary but rarely practiced in chronic disease settings. And research on human-AI interaction in chronic care establishes that design principles must be grounded in evidence from both patient experience and expert clinical judgment---yet such dual-perspective evidence rarely exists. Our study addresses all three simultaneously: we use \revise{participant reported patient queries} to evaluate AI outputs across multiple dimensions with physician judgment, then analyze where patient and clinician perspectives converge and where they diverge \revise{through triangulation across Study~1 and Study~2}.

\section{OBSERVATIONAL STUDY OF PATIENT USE AND ATTITUDES}
This formative study examines how Chinese patients with T2DM interact with generative AI in daily self-management, characterizing question types, usage patterns, and attitudes that informed the evaluation design in Study~2.

\subsection{Method}
\subsubsection{Participants}
We recruited participants through online platforms (Xiaohongshu, Bilibili, Baidu Tieba) and offline outreach across four Northern Chinese provinces (Henan, Hebei, Shanxi, Shandong). Eligibility required participants to be aged 18 or older, clinically diagnosed with T2DM, and to have prior experience using generative AI for diabetes-related information seeking.

A total of 21 participants were enrolled (\revise{11 men, 10 women}; age: 33--65, $M=47.76$, $SD=8.14$; T2DM duration: 1--23 years, $M=8.10$, $SD=5.52$; 12 rural, 9 urban; education: 4 bachelor's, 7 high school, 3 middle school, 2 primary school, 5 no formal education; Table~\ref{tab:obs_participants}). IRB approval was obtained from [Anonymous University]; participants provided written informed consent and received 30 RMB compensation.

\begin{table}[htbp]
\centering
\caption{Demographic characteristics of participants in the observational study (N=21).}
\label{tab:obs_participants}
\begin{tabular}{cccccc}
\toprule
ID & Age & Years with T2DM & Gender & Residence & Education \\
\midrule
1  & 44 & 5  & \revise{Woman} & Urban & High school \\
2  & 45 & 5  & \revise{Man} & Urban & High school \\
3  & 54 & 4  & \revise{Man} & Rural & Middle school \\
4  & 55 & 12 & \revise{Man} & Urban & Middle school \\
5  & 56 & 13 & \revise{Woman} & Rural & Primary school \\
6  & 65 & 23 & \revise{Woman} & Urban & Primary school \\
7  & 57 & 17 & \revise{Man} & Rural & None \\
8  & 56 & 11 & \revise{Woman} & Urban & High school \\
9  & 45 & 11 & \revise{Man} & Rural & High school \\
10 & 38 & 3  & \revise{Woman} & Urban & Bachelor \\
11 & 36 & 3  & \revise{Man} & Urban & Bachelor \\
12 & 39 & 2  & \revise{Woman} & Urban & High school \\
13 & 33 & 4  & \revise{Man} & Rural & Bachelor \\
14 & 45 & 11 & \revise{Man} & Rural & High school \\
15 & 56 & 12 & \revise{Woman} & Urban & None \\
16 & 52 & 10 & \revise{Woman} & Rural & None \\
17 & 49 & 6  & \revise{Man} & Rural & High school \\
18 & 41 & 1  & \revise{Man} & Rural & Bachelor \\
19 & 46 & 3  & \revise{Woman} & Urban & Middle school \\
20 & 44 & 7  & \revise{Man} & Rural & None \\
21 & 47 & 7  & \revise{Woman} & Rural & None \\
\bottomrule
\end{tabular}
\end{table}

\subsubsection{Procedure}
The study comprised three components. First, participants completed two structured questionnaires: an AI Attitude Questionnaire (10 items, 10-point Likert scale; Appendix~A) and an AI Usage Questionnaire capturing platform preferences and interaction scenarios (Appendix~B). Open-ended fields invited participants to document questions previously posed to AI tools \revise{by transcribing query text or paraphrasing from memory}, yielding a corpus of \textbf{784 \revise{participant reported diabetes related queries}} that formed the basis of Study~1's thematic analysis and Study~2's question set. \revise{We treat this corpus as evidence of patient reported information needs rather than as a complete unedited interaction log.} Second, 35--55 minute semi-structured interviews (Appendix~C) explored usage contexts, perceived benefits and risks, and expectations for future AI capabilities. Interviews were audio-recorded with consent and transcribed within 24 hours.

\subsubsection{Data Analysis}
Quantitative survey data were analyzed descriptively. Qualitative data, including \revise{patient submitted} queries and interview transcripts, were analyzed using \revise{codebook based thematic analysis}~\cite{braun2006using}, supported by NVivo. Two researchers independently coded an initial subset to construct a shared codebook, and \revise{coding agreement} exceeded 85\%, with discrepancies resolved by consensus. Through iterative inductive and deductive coding, we identified \textbf{seven recurring categories} of patient informational need, \textit{\revise{Factual Knowledge}, Diet Management, Sports Advice, Medication Guide, Medication Interpretation, Complications Related}, and \textit{Life and Psychology}, which structured both Study~1's analysis and Study~2's question corpus.

\subsection{Results}

\subsubsection{AI Usage Patterns and Question Themes}

\begin{figure}[htbp]
    \centering
    \includegraphics[width=0.8\linewidth]{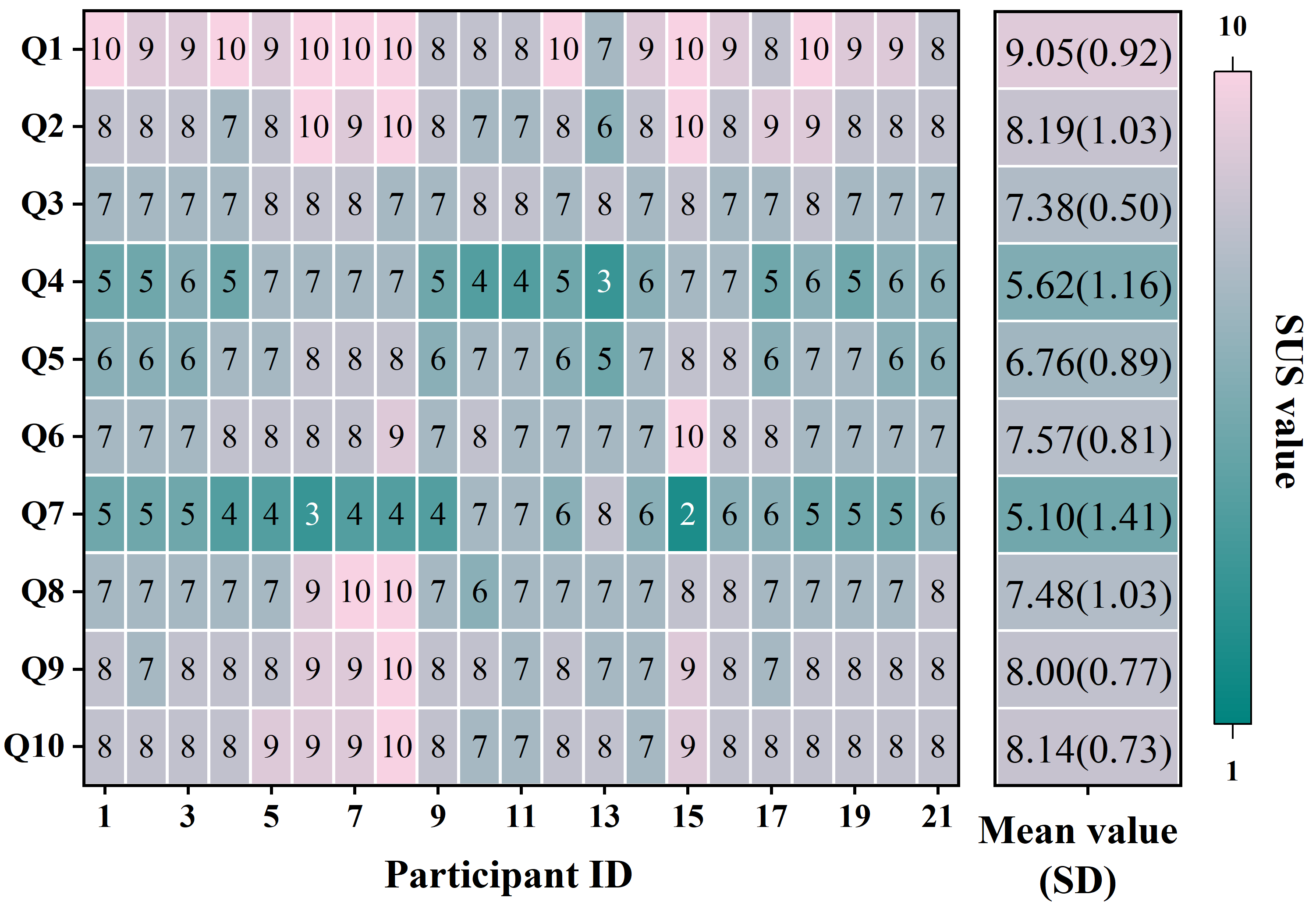}
    \caption{Heatmap of System Usability Scale (SUS)-style ratings from Study~1 participants.
    Each cell shows one participant's score (1--10) on 10 items, with color encoding magnitude. Right column displays means (SDs).}
    \label{fig:study1_sus_heatmap}
\end{figure}

Figure~\ref{fig:study1_sus_heatmap} presents participants' attitude ratings. Scores were high for perceived convenience (Q1: $M=9.05$, $SD=0.92$), accessibility (Q2: $M=8.19$), and general satisfaction (Q10: $M=8.14$), but substantially lower for psychological support (Q7: $M=5.10$, $SD=1.41$) and personalization (Q4: $M=5.62$), indicating perceived limitations in AI's human-centered capabilities.

Textual analysis of the 784 \revise{participant reported} queries revealed diverse informational needs across seven domains. \revise{The curated Study~2 question set included 14 Diet Management items, 11 Life and Psychology items, and 9 Factual Knowledge items. These item counts refer to representative question types selected for physician evaluation rather than raw category frequencies in the full corpus.} Diet questions ranged from glycemic index to cultural food pairing, reflecting expectations for behavioral guidance calibrated to daily life rather than generic information. \revise{Factual Knowledge} questions, such as ``Can T2DM be cured?'' or ``What is HbA1c?'', positioned AI as a first point of contact for illness literacy, filling gaps left by time-constrained clinical consultations.

\textit{Medication Guide} and \textit{Medication Interpretation} questions carried high clinical stakes: patients asked about drug effects and dosing uncertainty without awareness of comorbidity risks or drug interactions---a gap with direct safety implications. \textit{Life and Psychology} questions revealed a qualitatively different register of need---anxiety, social stigma, and practical logistics such as insulin storage required emotional attunement rather than factual retrieval. Across all categories, a persistent pattern emerged: patients sought not population-level generalizations but contextual judgment tailored to their individual circumstances.

These patterns \revise{informed} five dimensions warranting clinical evaluation: guideline alignment (\textbf{Accuracy}), absence of harmful advice (\textbf{Safety}), accessible language (\textbf{Clarity}), information completeness (\textbf{Integrity}), and actionable guidance (\textbf{Action Orientation}), operationalized in Study~2.

\subsubsection{Patient Attitudes Toward AI and Perceived Boundaries}

Patients expressed cautious optimism, valuing AI for routine queries and preparation for clinical encounters. A recurring \textit{pre-visit primer} role emerged: \textit{``After chatting with AI, I know what exactly to ask my doctor---it saves time.''} This function was especially valued in China's overburdened clinical system, where limited consultation time makes patient-initiated health literacy consequential for the quality of doctor--patient interaction.

Participants maintained clear role boundaries. Physician authority remained the ultimate decision-making locus: \textit{``If AI says one thing and my doctor says another, I will always follow the doctor.''} Two friction points recurred consistently. First, population-level advice felt inadequate or misleading when not adjusted for individual comorbidities, cultural dietary practices, or economic constraints. Second, AI's communicative register proved ill-suited to the emotional demands of chronic illness: responses to anxiety or diabetes burnout were consistently described as robotic and impersonal.

\subsubsection{Expectations for Future AI Health Assistants}

Participants converged on three expectations: \textit{contextual personalization}---responses that account for individual comorbidities and constraints, not fixed templates (P3: \textit{``I hope AI can truly understand my condition, not just repeat a fixed template''}); \textit{source transparency}---traceable provenance to guidelines or verified sources (P13: \textit{``If AI could show where its advice comes from---like official guidelines---I would feel more reassured''}); and \textit{ecosystem integration}---connection to wearables, physiological sensors, and optional physician review pathways. These expectations directly motivate the design directions developed in the Discussion.

\section{EVALUATION STUDY OF AI-GENERATED HEALTH INFORMATION BY PHYSICIANS}
Building on the patient usage data from Study~1, we conducted a structured evaluation study to assess the quality of AI-generated health information for T2DM management from a clinical perspective. Seven endocrinologists independently evaluated responses from four AI systems using a \revise{structured five dimensional physician rating rubric}, followed by in-depth interviews to surface their evaluative reasoning and design expectations.

\subsection{Participants}

We recruited participating physicians through online outreach on major Chinese platforms, including Xiaohongshu, Bilibili, and Baidu Tieba. To be eligible, physicians were required to meet \revise{four criteria. They had to} (1) hold a valid medical license, (2) be currently practicing in endocrinology or have long-term experience treating T2DM, (3) have a minimum of five years of clinical experience in T2DM care, and (4) hold the professional rank of attending physician or higher. All participants were required to provide written informed consent. Physicians were excluded if they were directly involved in the development of AI tools for diabetes care or held financial interests in related companies, to avoid conflicts of interest.

A total of seven physicians participated in this evaluation study. Their ages ranged from 34 to 62 years ($M=48.86$, $SD=11.57$), with years of clinical practice ranging from 5 to 33 years ($M=19.86$, $SD=11.60$). All participants held a PhD degree. Five physicians specialized in endocrinology or metabolic disorders, one in rehabilitation medicine, and one in cardiology. Five practiced in urban settings and two in rural areas. Table~\ref{tab:physicians} summarizes the demographic characteristics of the participating physicians.

This study was approved by the Institutional Review Board (IRB) of [Anonymous University]. Participants provided written informed consent prior to participation. They retained the right to withdraw from the study or request data deletion at any time. All data were anonymized during collection and analysis. Each physician received a compensation of 100 RMB for their participation.

\begin{table}[htbp]
\centering
\caption{Demographic characteristics of participating physicians (N=7).}
\label{tab:physicians}
\begin{tabular}{ccccccc}
\toprule
ID & Age & Years of Practice & Department & Gender & Residence & Degree \\
\midrule
1 & 57 & 28 & Endocrinology & \revise{Man} & Urban & PhD \\
2 & 62 & 33 & Endocrinology & \revise{Man} & Urban & PhD \\
3 & 57 & 29 & Endocrinology & \revise{Man} & Urban & PhD \\
4 & 55 & 25 & Rehabilitation & \revise{Woman} & Rural & PhD \\
5 & 34 & 5  & Endocrinology and Metabolism & \revise{Man} & Urban & PhD \\
6 & 42 & 13 & Cardiology & \revise{Woman} & Urban & PhD \\
7 & 35 & 6  & Endocrinology & \revise{Woman} & Rural & PhD \\
\bottomrule
\end{tabular}
\end{table}

\subsection{Procedures}
The evaluation study consisted of three sequential phases: (1) physician-led selection of patient-generated questions, (2) expert evaluation of AI responses, and (3) post-evaluation interviews and reflections.

\paragraph{\textbf{Phase 1: Question Selection}}
In the first phase, the physician team reviewed all 784 \revise{participant reported} questions collected from patients in Study~1. \revise{The 784 questions were not all entered into the AI systems. Instead, physicians and researchers curated a 66 question evaluation set.} Three criteria guided curation of a representative subset: (1) frequency of occurrence, (2) clinical significance (i.e., relevance to core management decisions), and (3) potential risk if answered incorrectly. Through iterative discussion and consensus, the team finalized a set that balanced high-frequency queries with low-frequency but high-risk items.

Questions were categorized into seven domains: \textit{\revise{Factual Knowledge}, Diet Management, Sports Advice, Medication Guide, Medication Interpretation, Complications Related}, and \textit{Life and Psychology}. This taxonomy extends the standard self-management pillars (education, diet, exercise, medication, monitoring) to include psychosocial dimensions in line with the biopsychosocial model of care. The final question set is presented in Appendix~D.

\paragraph{\textbf{Phase 2: AI Output Evaluation}}

Each question was entered into four mainstream generative AI platforms: GPT-4-turbo \cite{openai_chatgpt_2025}, DeepSeek-R1 \cite{deepseek_2025}, Kimi K2 \cite{kimi_2025}, and ERNIE Bot 4.5 \cite{erniebot_2025}. To ensure independence and comparability, a new session was initiated for each query and memory or context-continuation features were disabled where possible. All responses were saved, anonymized, and archived by question ID and platform type.

Each physician independently evaluated the AI outputs using a five-dimensional rubric with a maximum total score of 120 points, \revise{developed through an iterative process grounded in Study~1 findings and physician priorities} (Section~3). \revise{Researchers first summarized patient query categories and recurrent response concerns from Study~1. The physician team then reviewed these categories and identified response qualities that would determine whether an AI answer was appropriate for patient facing use. The rubric was used as a structured expert evaluation tool, not as a psychometrically validated scale.} The rubric assessed each dimension as follows:

\begin{itemize}
  \item \textbf{Accuracy (30)}: Alignment with current clinical guidelines and evidence-based practice.
  \item \textbf{Safety (30)}: Absence of misleading or potentially harmful advice (e.g., unsupervised drug discontinuation, unverified therapies).
  \item \textbf{Clarity (20)}: Use of accessible, non-technical language and clear explanatory structure.
  \item \textbf{Integrity (20)}: Inclusion of all key components \revise{and} avoidance of partial or fragmented information.
  \item \textbf{Action Orientation (20)}: Provision of concrete, practical suggestions for health self-management.
\end{itemize}

The differential weighting reflects clinical priority in patient-facing health communication. \textit{Accuracy} and \textit{Safety} carry the highest weight (30 points each) because errors in these dimensions carry direct clinical risk: a factually incorrect or potentially harmful recommendation cannot be offset by superior linguistic quality or completeness. \textit{Clarity}, \textit{Integrity}, and \textit{Action Orientation} (20 points each) are weighted equally, reflecting their comparable contribution to communicative effectiveness in patient education. This weighting scheme was deliberated and agreed upon by the physician team prior to evaluation.

Physicians could optionally add qualitative comments to highlight strengths, deficiencies, or safety concerns. All scores and written feedback were compiled and \revise{reviewed for completeness}, and served as the basis for Phase~3.

\paragraph{\textbf{Phase 3: Expert Interviews and Reflective Analysis}}

To gain deeper insight into physicians' evaluative reasoning, we conducted semi-structured interviews following the scoring phase (full protocol in Appendix~E). Each 30--60 minute session covered three areas: overall impressions of AI utility and reliability, dimension- and domain-specific reasoning about the five evaluation criteria and seven question categories, and forward-looking expectations for AI integration in clinical workflows. Interviewers used structured prompts to elicit concrete examples (e.g., ``Can you recall an example where an AI response felt unsafe?''). All sessions were audio-recorded with consent, transcribed within 24 hours, and manually verified to ensure fidelity.

\subsection{Data Analysis}

Physician rating data (N = 7 raters $\times$ 4 models $\times$ 66 questions) were first cleaned and analyzed descriptively (means, SDs, IQRs). \revise{Given the small physician sample, inferential tests were treated as exploratory support for descriptive patterns.} To examine model-level differences, we conducted a one-way repeated-measures ANOVA~\cite{girden1992anova} with AI system as the within-subject factor. Mauchly's sphericity test was applied and Greenhouse--Geisser corrections used where violated, with post-hoc pairwise comparisons using Bonferroni correction. Effect sizes were reported as generalized eta squared ($\eta^2_G$). To examine interaction effects, two-way repeated-measures ANOVAs were conducted with Model $\times$ Dimension (5 levels) and Model $\times$ Category (7 levels) as within-subject factors, followed by simple effects tests. Score distributions were further visualized through boxplots, radar charts, and heatmaps.

The qualitative data from post-evaluation interviews were analyzed using \revise{codebook based thematic analysis} \cite{braun2006using}. Transcripts were manually coded in two iterative rounds: initial open coding was conducted independently by two researchers to generate a preliminary codebook, which was then refined through collaborative discussion. Codes were organized into dimension-aligned themes (e.g., perceptions of safety, critiques of clarity, emotional response gaps) and emergent design expectations (e.g., transparency, contextual reasoning, model orchestration). Representative quotes were selected to provide interpretive depth to quantitative trends. Thematic development was \revise{reviewed through peer debriefing and senior clinical input}.

\section{STUDY 2 RESULTS: PHYSICIAN EVALUATION OF AI-GENERATED HEALTH INFORMATION}
This section reports findings from Study~2, in which seven physicians evaluated AI-generated responses across four models and five dimensions. Findings from Study~1 (patient usage patterns and attitudes) are reported in Section~3.

\subsection{Quantitative Findings}

\subsubsection{Performance Variability Across Models}

\begin{figure}[htbp]
    \centering
    \includegraphics[width=0.5\linewidth]{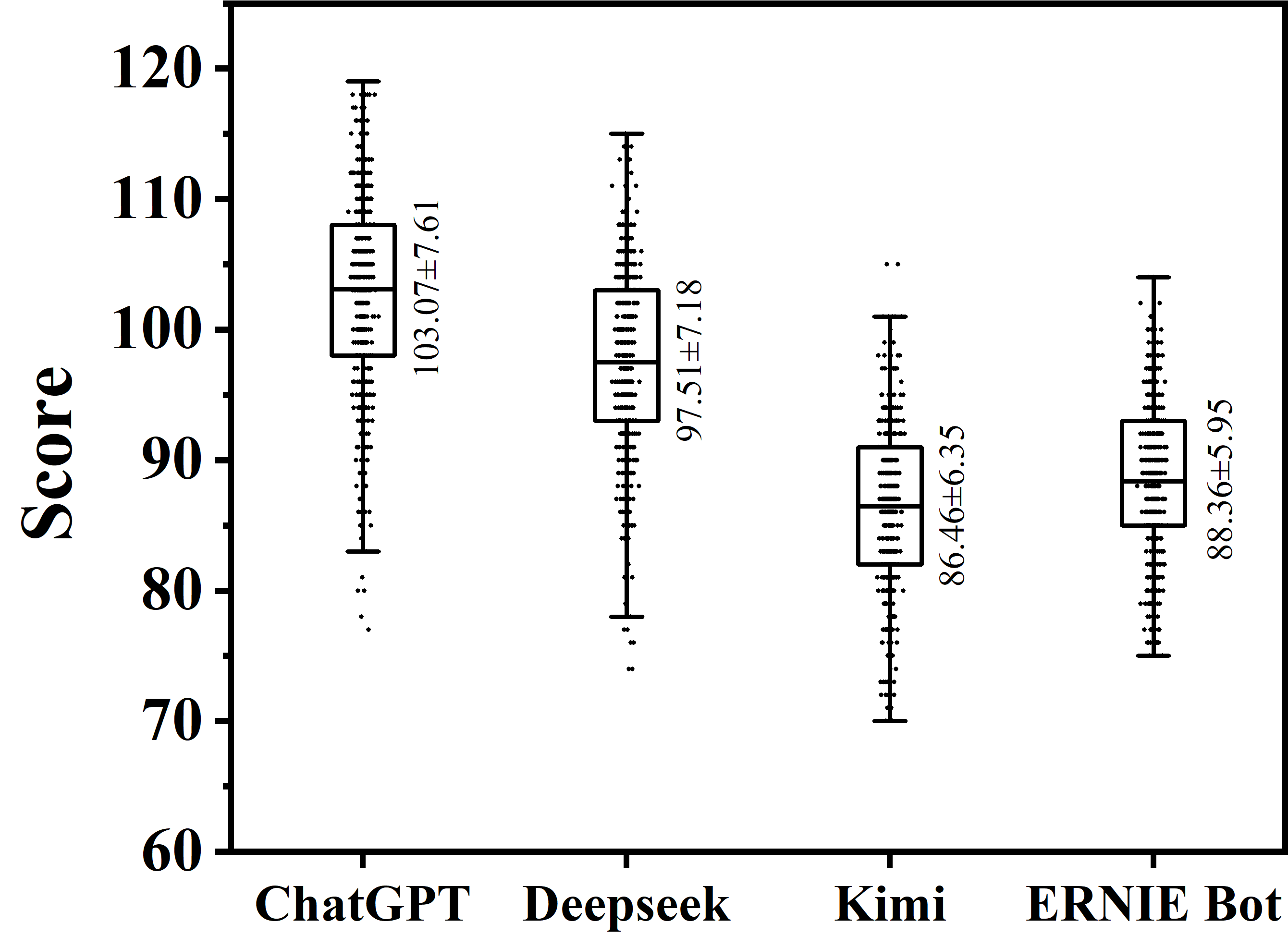}
    \caption{Overall quality evaluation of four AI models.
    Each box represents the distribution of aggregated scores across all evaluation dimensions and question types for one AI system.
    The figure provides an overall comparison of model quality.}
    \label{fig:overall_quality}
\end{figure}

To assess the overall quality of health information generated by different AI models, we first aggregated physician ratings across all questions and evaluation dimensions. As shown in Figure~\ref{fig:overall_quality}, ChatGPT achieved the highest mean score ($M=103.07$, $SD=7.61$), followed by DeepSeek ($M=97.51$, $SD=7.18$). ERNIE Bot ($M=88.36$, $SD=5.95$) and Kimi ($M=86.46$, $SD=6.35$) received considerably lower ratings.

Notably, ChatGPT and DeepSeek also exhibited greater score variability (wider IQRs and longer whiskers), while Kimi and ERNIE Bot showed smaller dispersion. This pattern \revise{suggests a possible} asymmetry in this dataset: the higher-performing models \revise{appeared} more sensitive to question type---excelling in favorable domains while declining more steeply in challenging ones---whereas the lower-performing models maintain a more uniform, if consistently lower, baseline across question types. This variability structure motivates the category-level analysis in the following subsection.

\revise{Exploratory statistical analysis was consistent with} these differences. A repeated-measures ANOVA \revise{showed} a significant effect of model type on overall quality, $F(3,18) = 2315.46$, $p < .001$, $\eta^2_G = 0.72$, indicating a large effect. The high F-value \revise{is consistent with strong agreement} among physicians in their relative model rankings, producing minimal within-model variance relative to the substantial between-model differences. Post-hoc tests showed that ChatGPT significantly outperformed all three alternatives ($p < .001$ for each), and DeepSeek also scored higher than both Kimi and ERNIE Bot. No reliable difference was observed between the latter two.

\begin{figure}[htbp]
    \centering
    \begin{subfigure}[t]{0.48\linewidth}
        \centering
        \includegraphics[width=\linewidth]{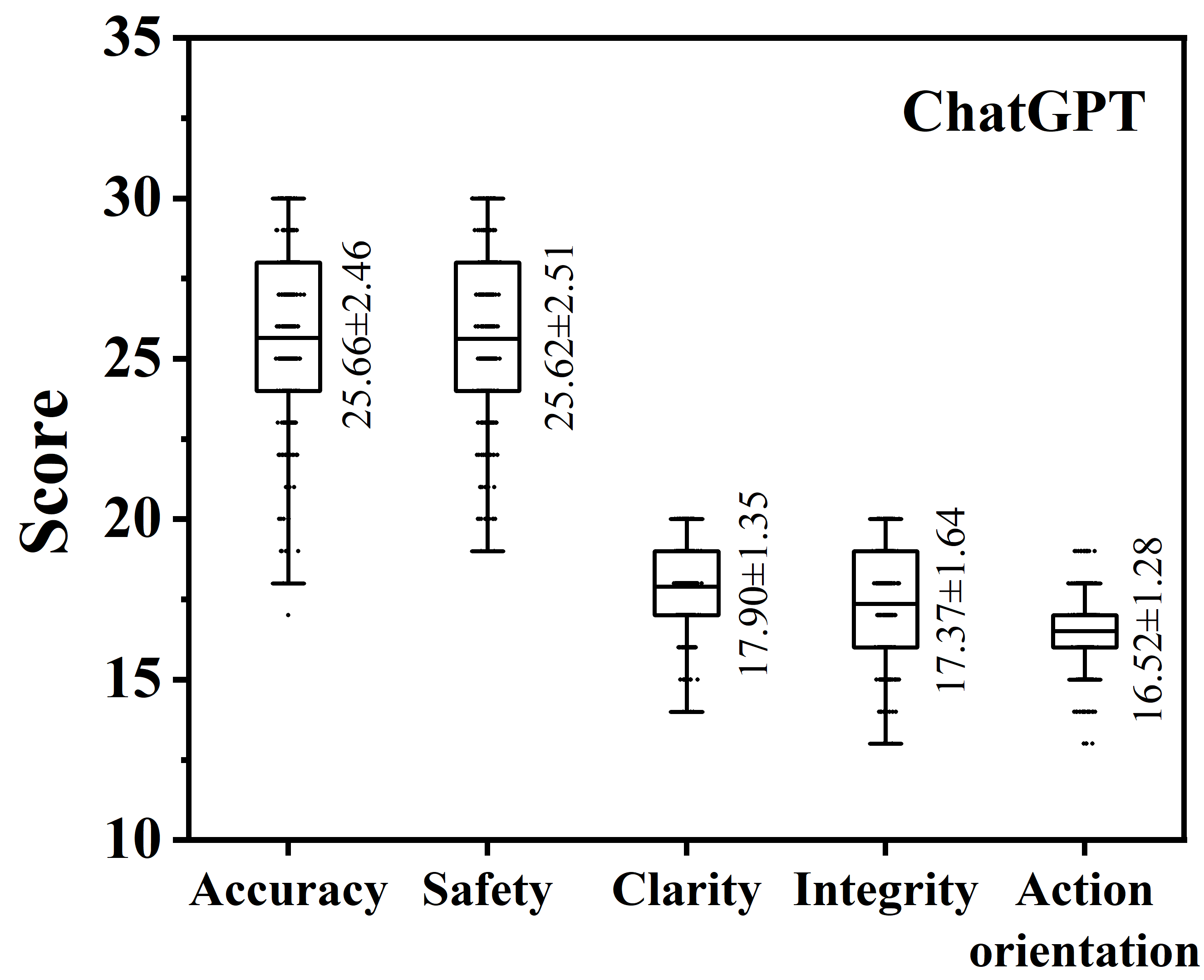}
    \end{subfigure}
    \hfill
    \begin{subfigure}[t]{0.48\linewidth}
        \centering
        \includegraphics[width=\linewidth]{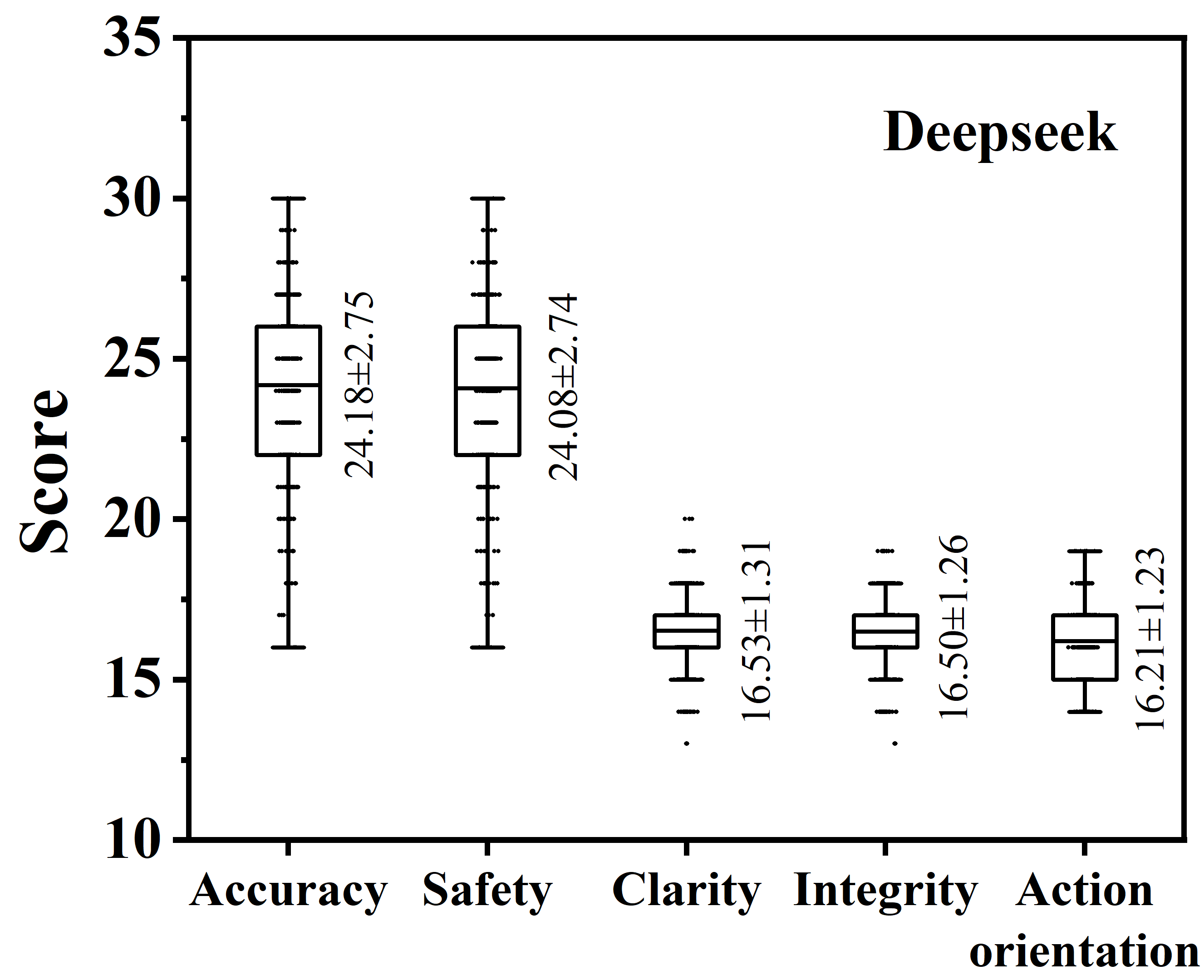}
    \end{subfigure}

    \vspace{1em}

    \begin{subfigure}[t]{0.48\linewidth}
        \centering
        \includegraphics[width=\linewidth]{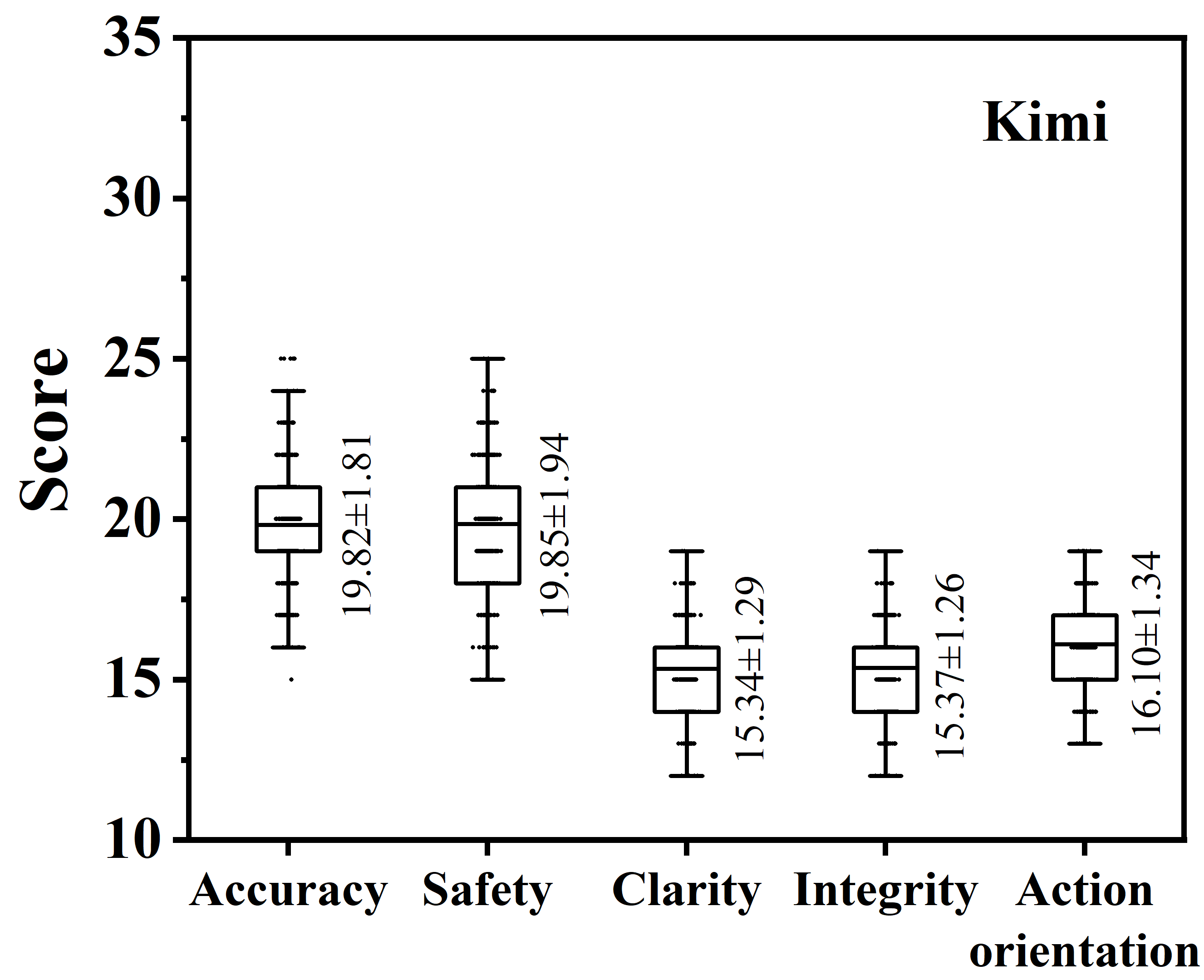}
    \end{subfigure}
    \hfill
    \begin{subfigure}[t]{0.48\linewidth}
        \centering
        \includegraphics[width=\linewidth]{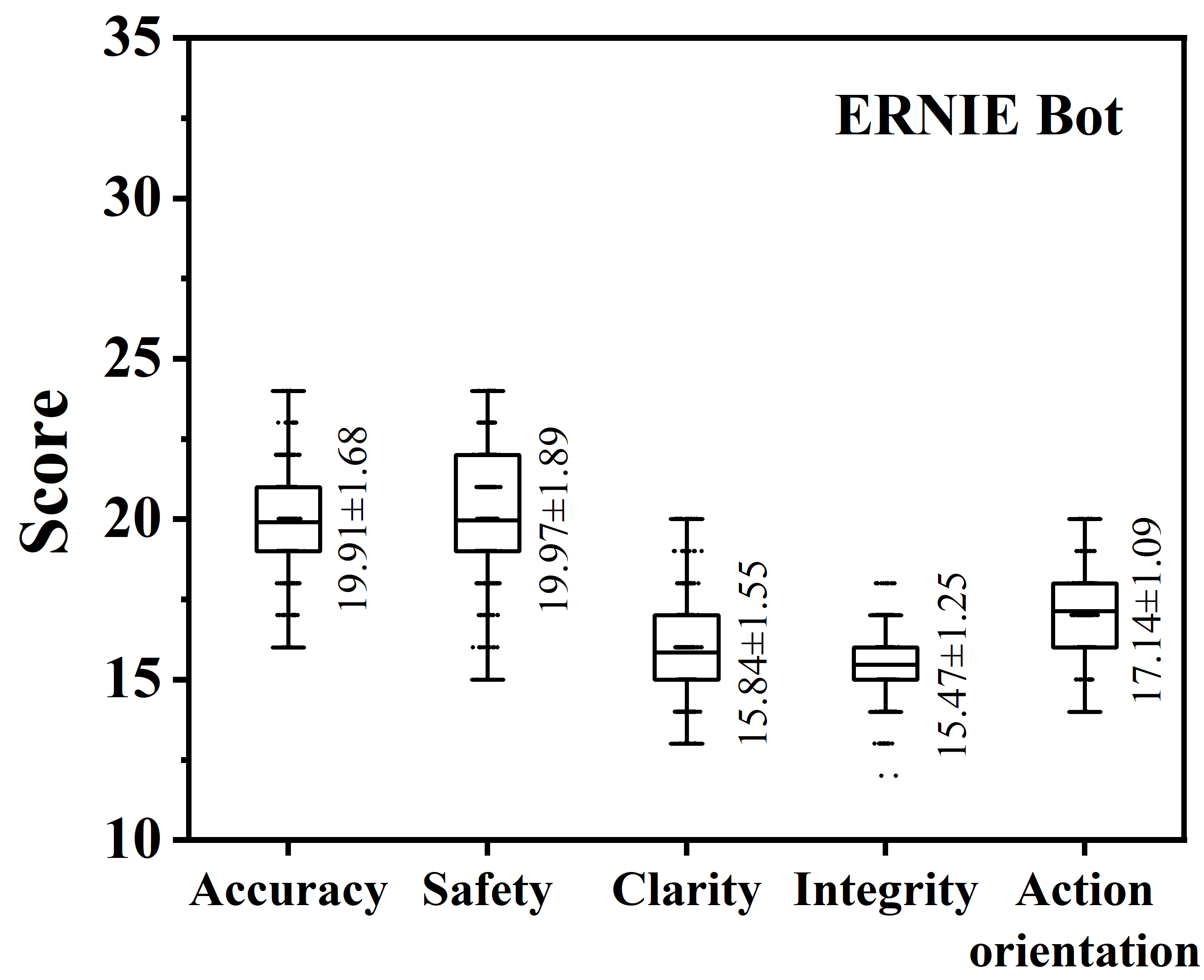}
    \end{subfigure}

    \caption{Dimension-wise quality evaluation of four AI models across five key criteria: Accuracy, Safety, Clarity, Integrity, and Action Orientation.
    Each subplot represents one AI model. Boxplots depict the distribution of scores across all evaluated questions under each criterion.}
    \label{fig:dimension_wise}
\end{figure}

\begin{figure}[htbp]
    \centering
    \includegraphics[width=0.6\linewidth]{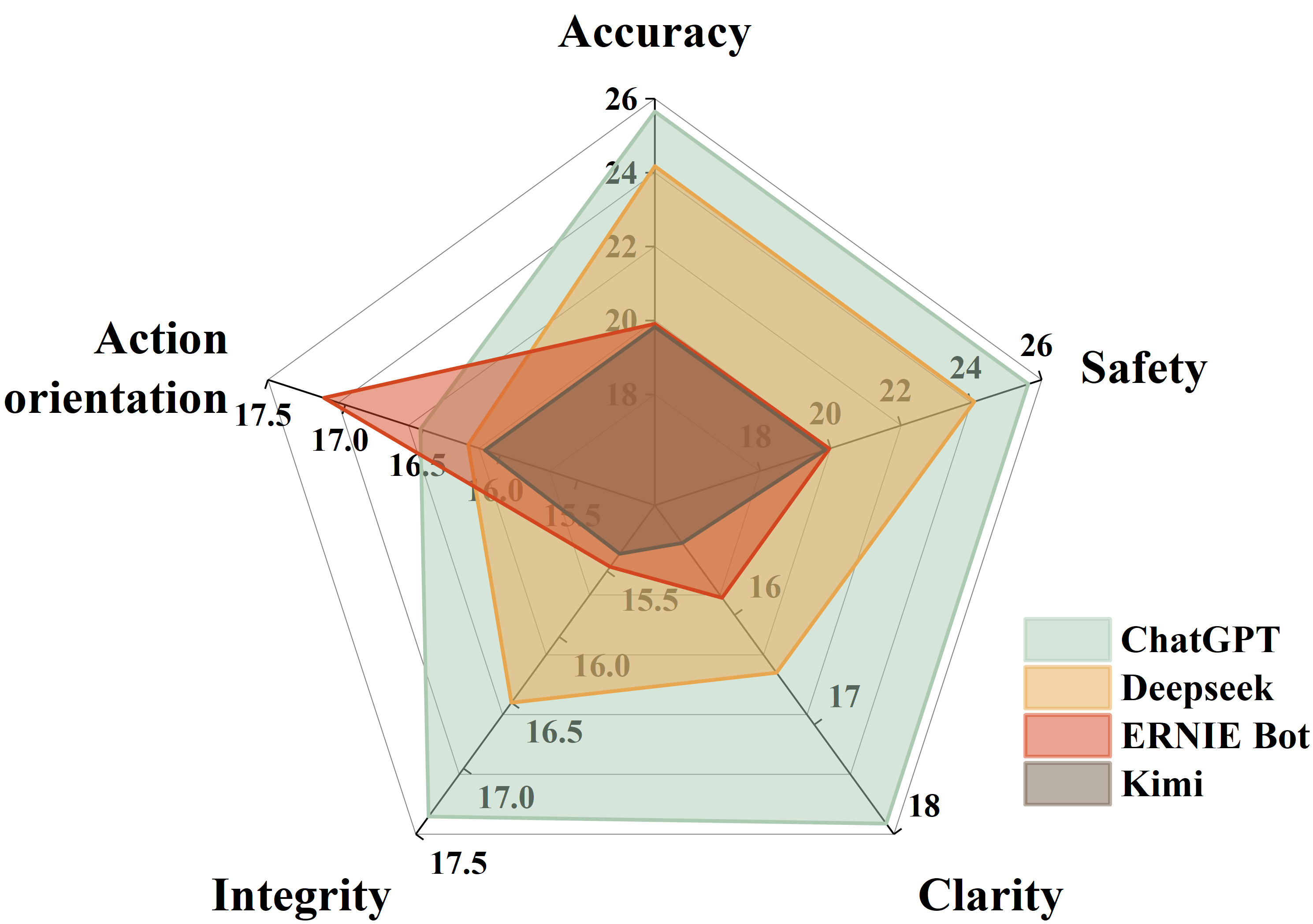}
    \caption{Dimension-wise comparison of AI models across five key quality dimensions:
    Accuracy, Safety, Clarity, Integrity, and Action Orientation.
    Each polygon represents one AI model's average score on the five dimensions.
    Larger enclosed areas indicate stronger overall performance across dimensions.}
    \label{fig:dimension_radar}
\end{figure}

To explore these differences further, we examined model behavior across five evaluation dimensions (Figures~\ref{fig:dimension_wise} and~\ref{fig:dimension_radar}). ChatGPT consistently achieved the highest scores across all five dimensions, including \textit{Accuracy} ($M=25.66$, $SD=2.46$) and \textit{Safety} ($M=25.62$, $SD=2.51$), suggesting reliable factual correctness and low-risk phrasing. Its \textit{Action Orientation} score ($M=16.52$), however, was its weakest dimension proportionally---a pattern consistent with physician observations that even high-performing models tend to accurately describe mechanisms without always translating them into patient-actionable steps.

DeepSeek's primary limitation was not in Clarity ($M=16.53$) or Integrity ($M=16.52$)---both comparable to ChatGPT's Clarity ($M=17.90$)---but rather a pronounced accuracy-safety cliff in medication-related categories, where its Accuracy dropped from approximately 27 in factual domains to approximately 20 in Medication Guide, a ~7-point decline more pronounced than ChatGPT's ~5-point drop. \revise{Physicians rated DeepSeek lower on tasks requiring contextual clinical reasoning.}

Among lower-performing models, Kimi scored lowest overall, with particular deficiencies in \textit{Clarity} ($M=15.34$) and \textit{Integrity} ($M=15.37$). ERNIE Bot showed a counterintuitive profile: despite its weakest overall score, it achieved the highest \textit{Action Orientation} across all four models ($M=17.14$)---surpassing even ChatGPT ($M=16.52$). Physicians noted that ERNIE Bot's directive, step-by-step output style produced high procedural scores, though its factual accuracy remained poor in safety-critical contexts---illustrating a core tension between directness and accuracy in AI health communication.

A two-way repeated-measures ANOVA revealed significant main effects of both Model ($F(3,18)=2315.46$, $p<.001$, $\eta^2=0.713$) and Dimension ($F(4,24)=4823.12$, $p<.001$, $\eta^2=0.922$), as well as a significant interaction ($F(12,72)=351.92$, $p<.001$, $\eta^2=0.606$), confirming that performance differences depend on both the model and the dimension of evaluation.

\subsubsection{Performance Variability Across Question Categories}

\begin{figure*}[htbp]
    \centering
    \includegraphics[width=0.5\linewidth]{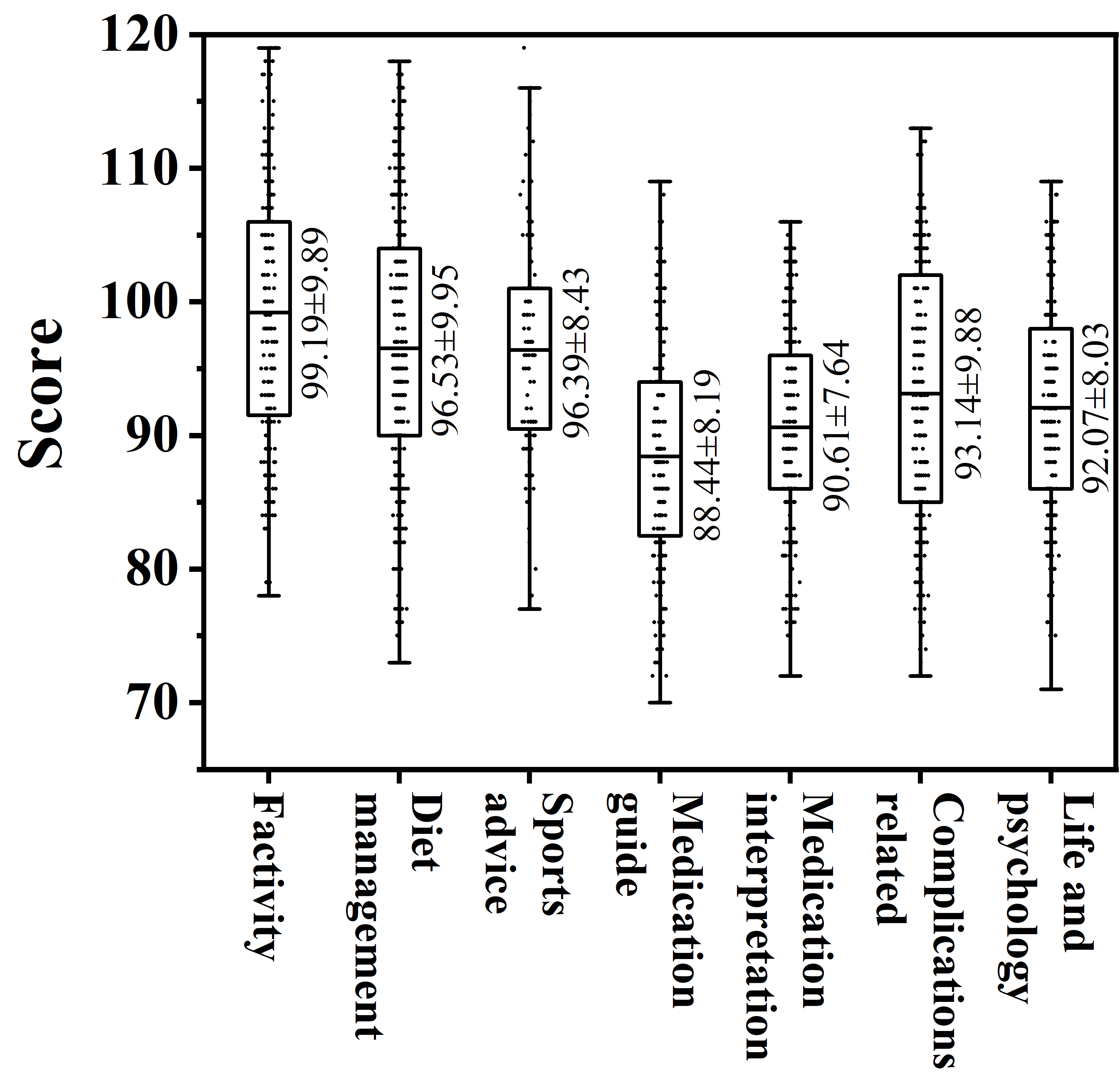}
    \caption{Overall physician evaluation of AI-generated information quality across seven question categories:
    \revise{Factual Knowledge}, Diet Management, Sports Advice, Medication Guide, Medication Interpretation, Complications Related, and Life and Psychology.
    Each box represents aggregated ratings across all four AI systems, illustrating cross-category variability and the relative difficulty of each question type.}
    \label{fig:overall_boxplot}
\end{figure*}

While model-level differences provide a coarse-grained view of performance, they do not reveal whether some types of patient queries are inherently more or less suitable for AI-assisted support. Figure~\ref{fig:overall_boxplot} presents the average evaluation scores across all four models for each of the seven question categories. The highest-performing domains were \textit{\revise{Factual Knowledge}} ($M=99.19$, $SD=9.89$), \textit{Diet Management} ($M=96.53$, $SD=9.95$), and \textit{Sports Advice} ($M=96.39$, $SD=8.43$). These topics involve standardized knowledge or general lifestyle recommendations, and AI responses were described as clear, reliable, and consistent with patient education materials or national guidelines.

By contrast, \textit{Medication Guide} ($M=88.44$) received the lowest scores overall---an approximately ten-point gap below \revise{Factual Knowledge}---and \textit{Medication Interpretation} ($M=90.61$) ranked second-lowest. These queries involve clinical judgment, such as dosage adjustment or interpretation of overlapping symptoms; physicians consistently flagged ambiguous phrasing and failure to differentiate between drug classes as recurring deficiencies. \textit{Complications Related} ($M=93.14$) and \textit{Life and Psychology} ($M=92.07$) reflected AI's difficulty with contextual reassurance and empathy: while systems often listed potential risks, they rarely balanced such warnings with evidence-based coping strategies.

\begin{figure*}[htbp]
    \centering
    \begin{subfigure}[t]{0.24\linewidth}
        \centering
        \includegraphics[width=\linewidth]{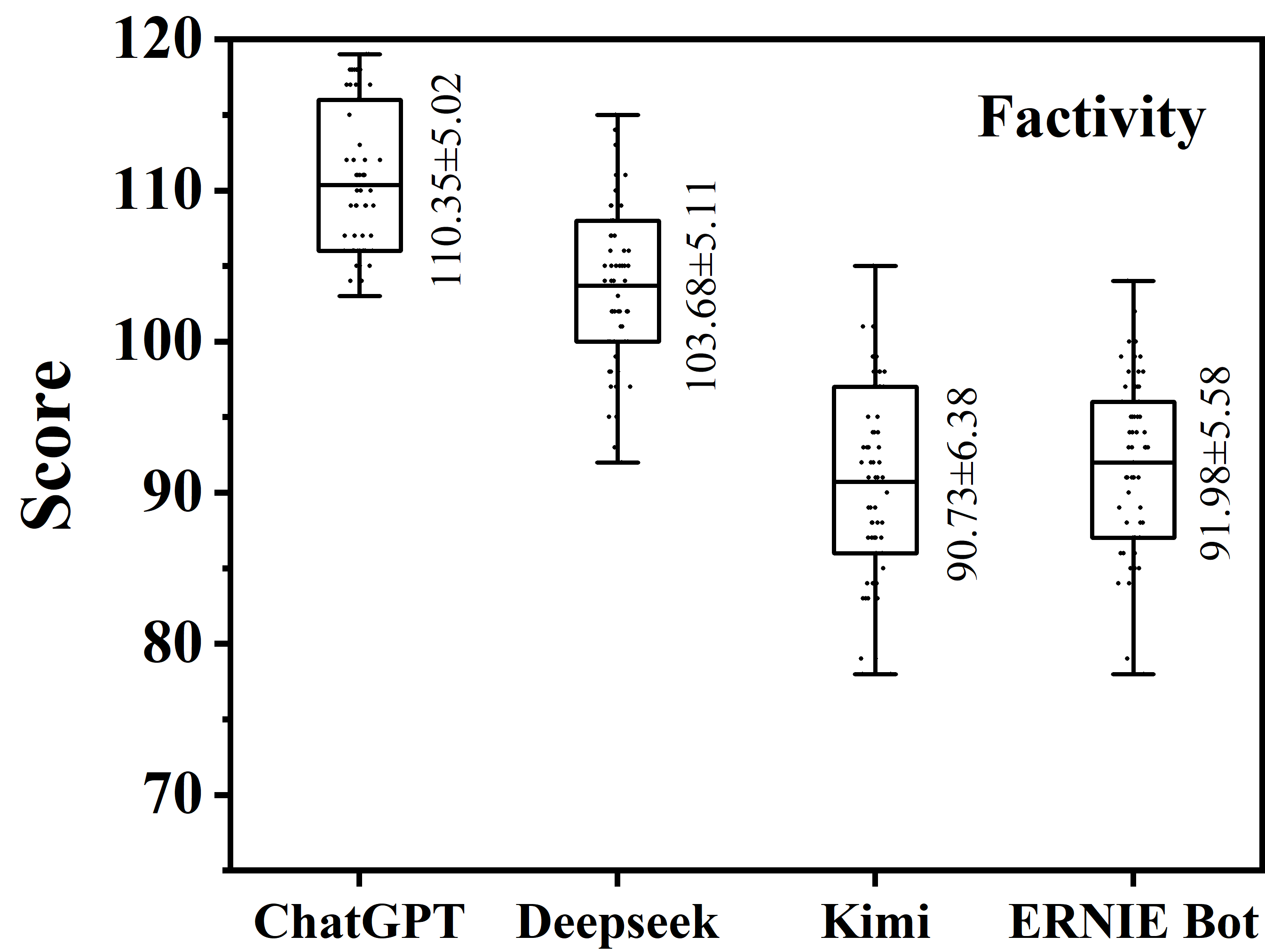}
    \end{subfigure}
    \hfill
    \begin{subfigure}[t]{0.24\linewidth}
        \centering
        \includegraphics[width=\linewidth]{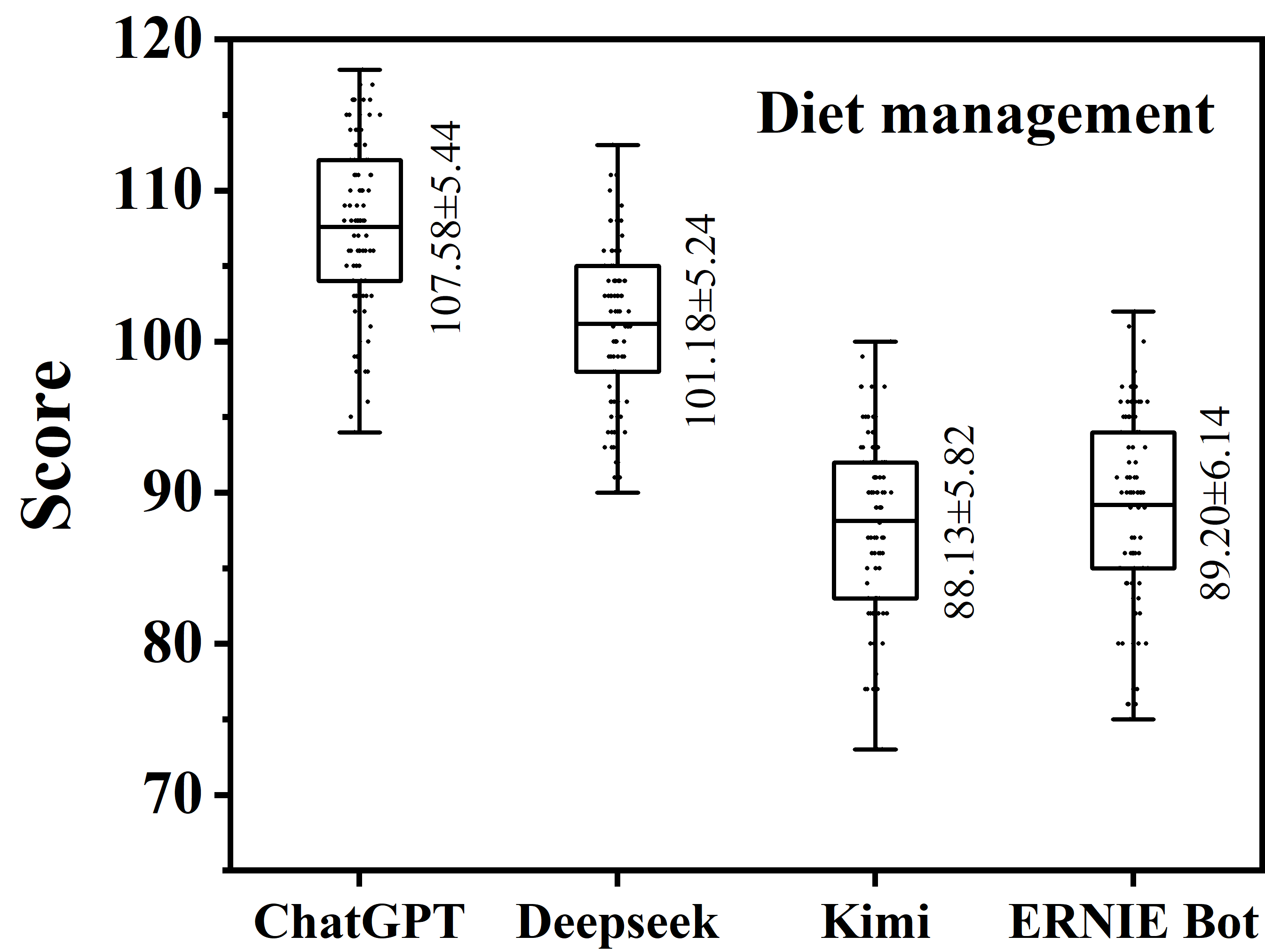}
    \end{subfigure}
    \hfill
    \begin{subfigure}[t]{0.24\linewidth}
        \centering
        \includegraphics[width=\linewidth]{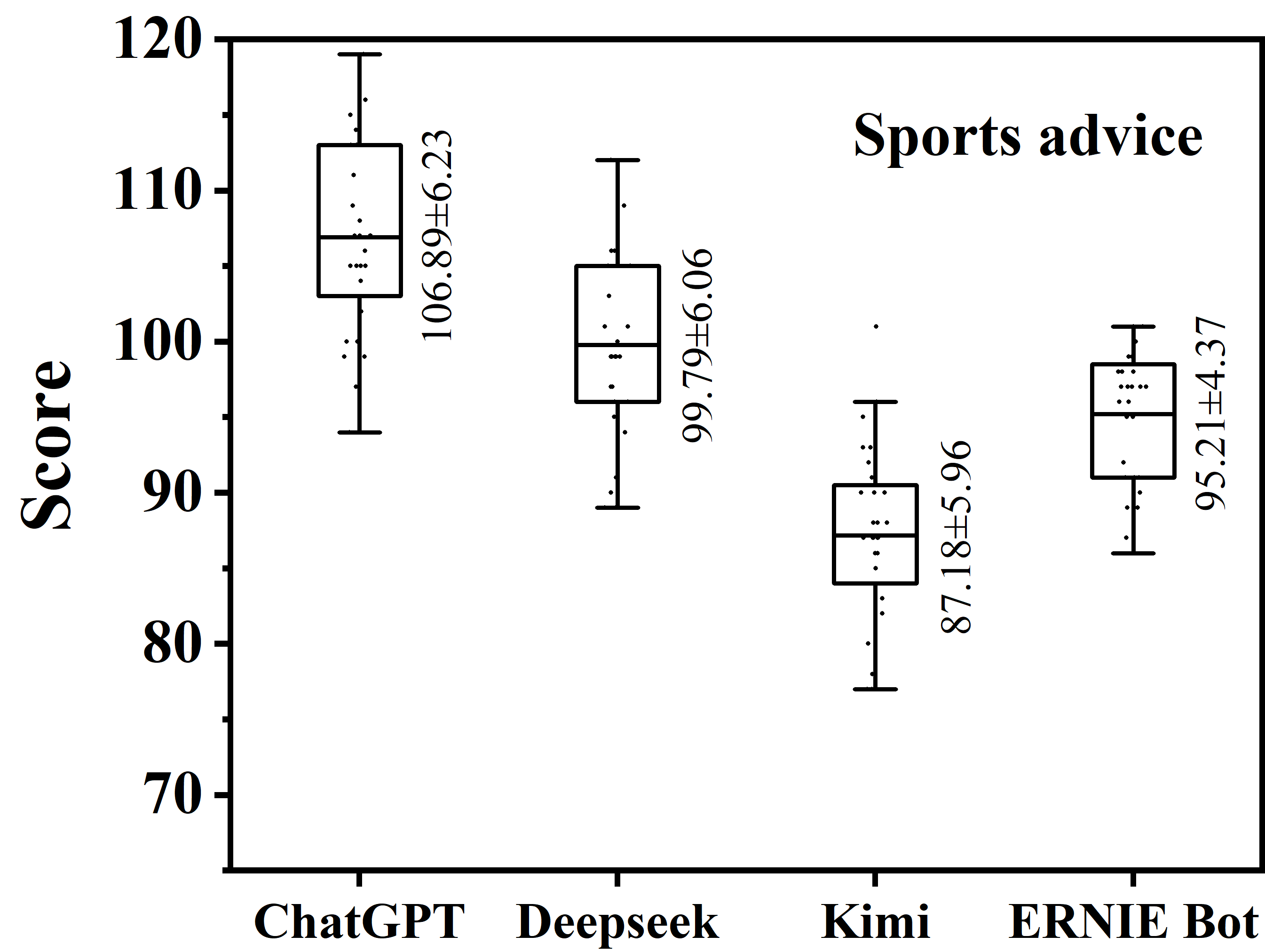}
    \end{subfigure}
    \hfill
    \begin{subfigure}[t]{0.24\linewidth}
        \centering
        \includegraphics[width=\linewidth]{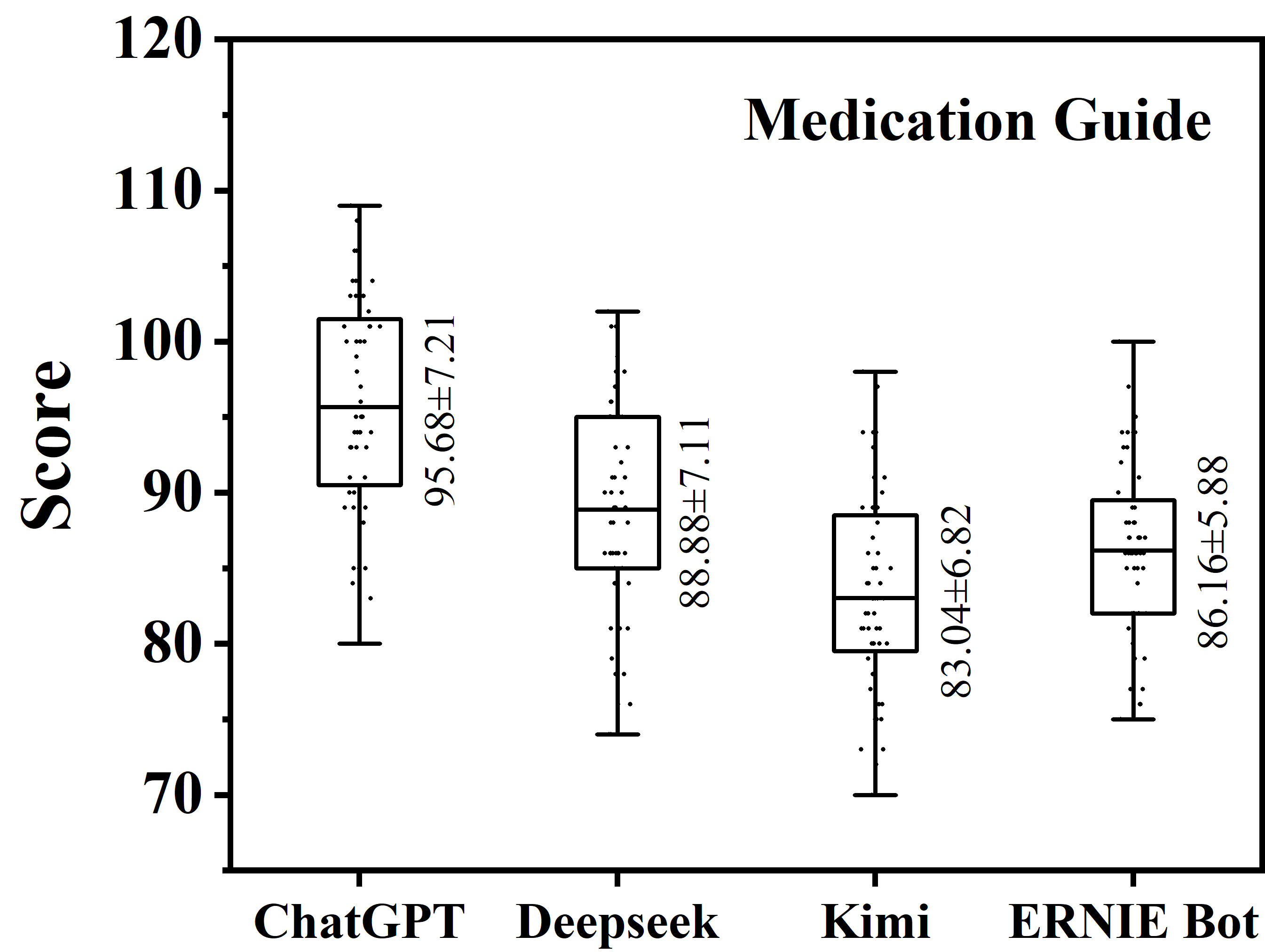}
    \end{subfigure}

    \vspace{0.8em}

    \begin{subfigure}[t]{0.24\linewidth}
        \centering
        \includegraphics[width=\linewidth]{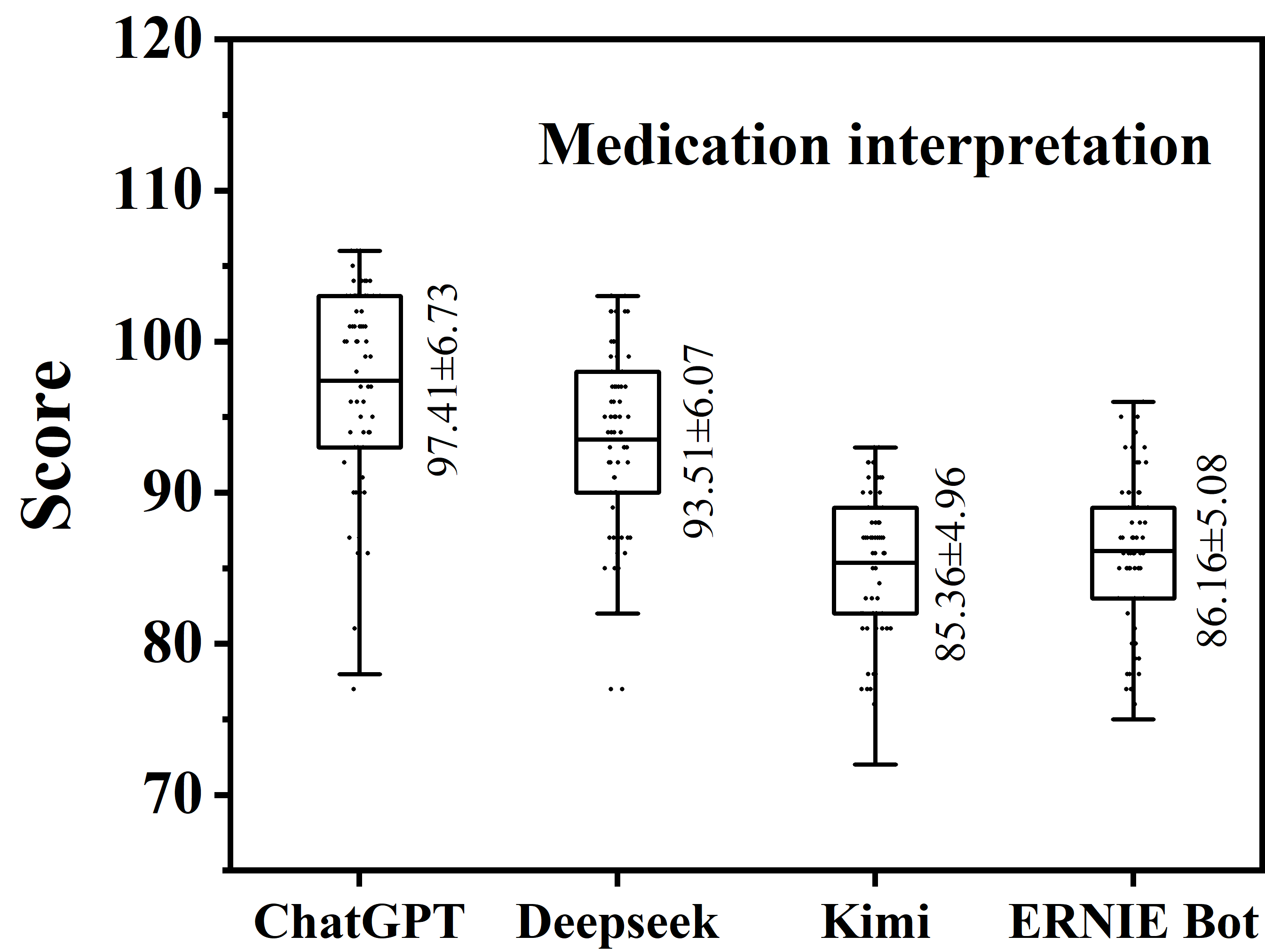}
    \end{subfigure}
    \hfill
    \begin{subfigure}[t]{0.24\linewidth}
        \centering
        \includegraphics[width=\linewidth]{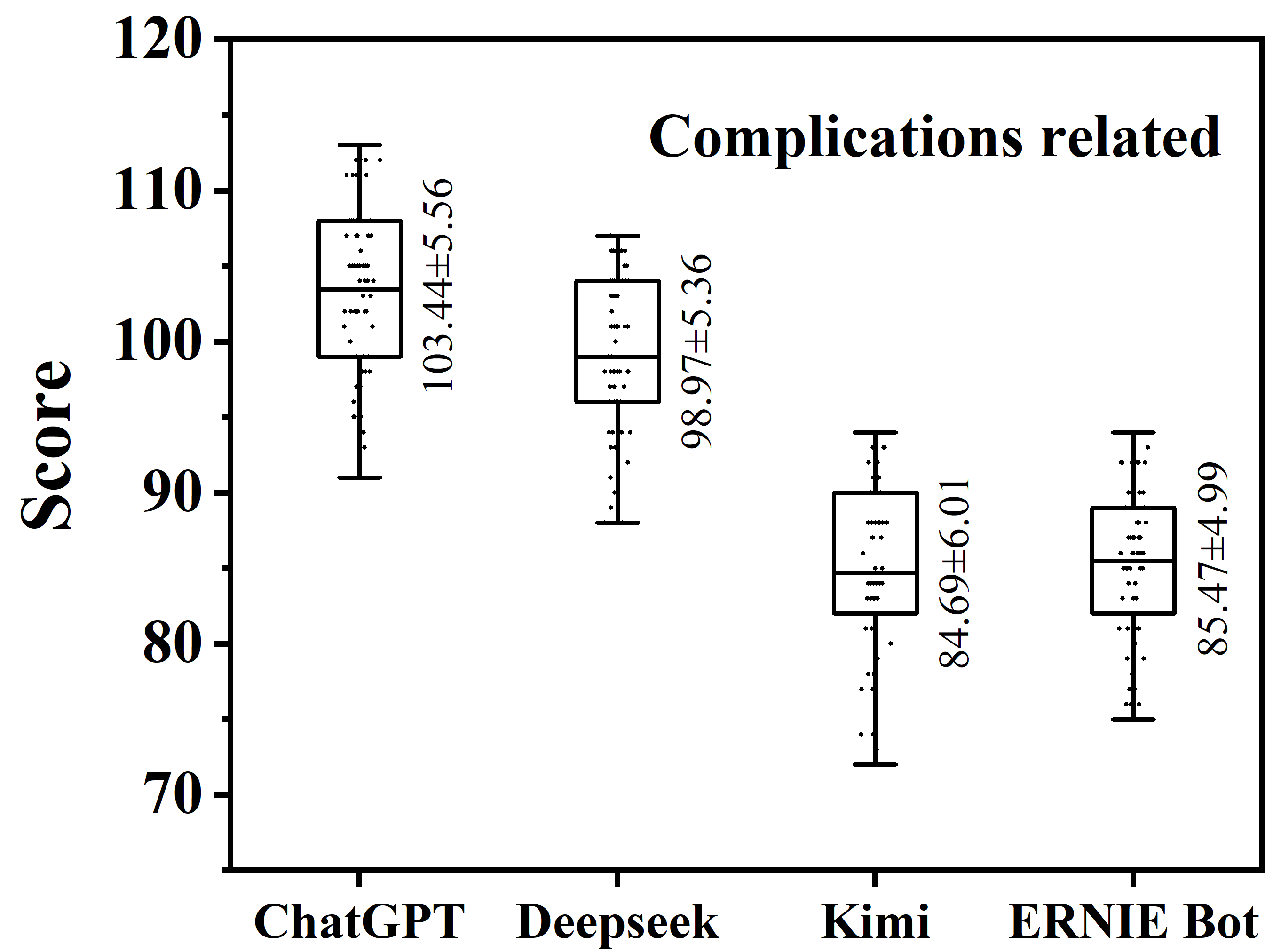}
    \end{subfigure}
    \hfill
    \begin{subfigure}[t]{0.24\linewidth}
        \centering
        \includegraphics[width=\linewidth]{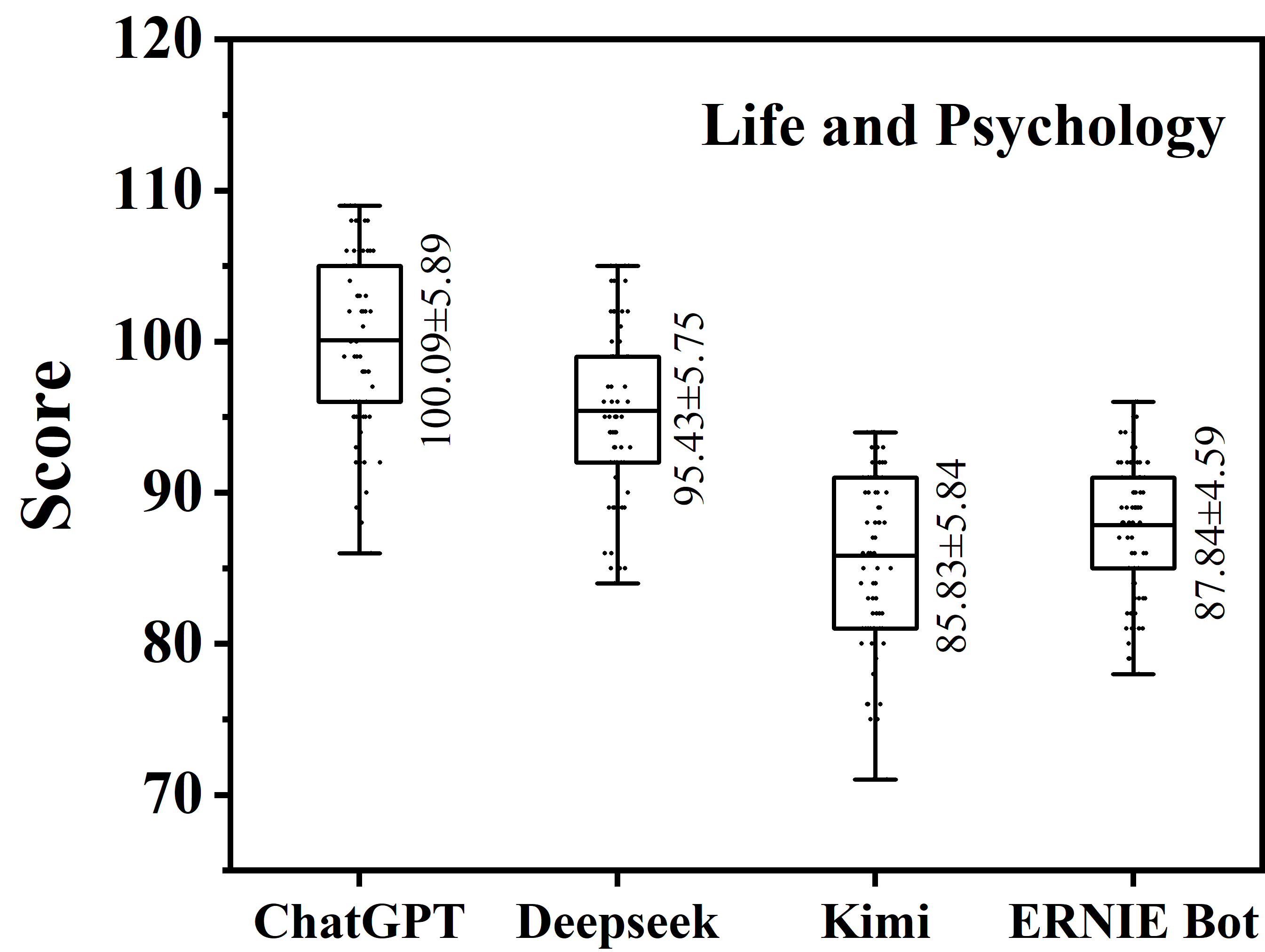}
    \end{subfigure}

    \caption{Dimension-level evaluation of AI-generated information quality across seven question categories.
    Each box represents aggregated physician ratings across all AI models for that category, reflecting the perceived suitability of AI responses for each topic area.}
    \label{fig:category_performance}
\end{figure*}

\begin{figure}[htbp]
    \centering
    \includegraphics[width=0.6\linewidth]{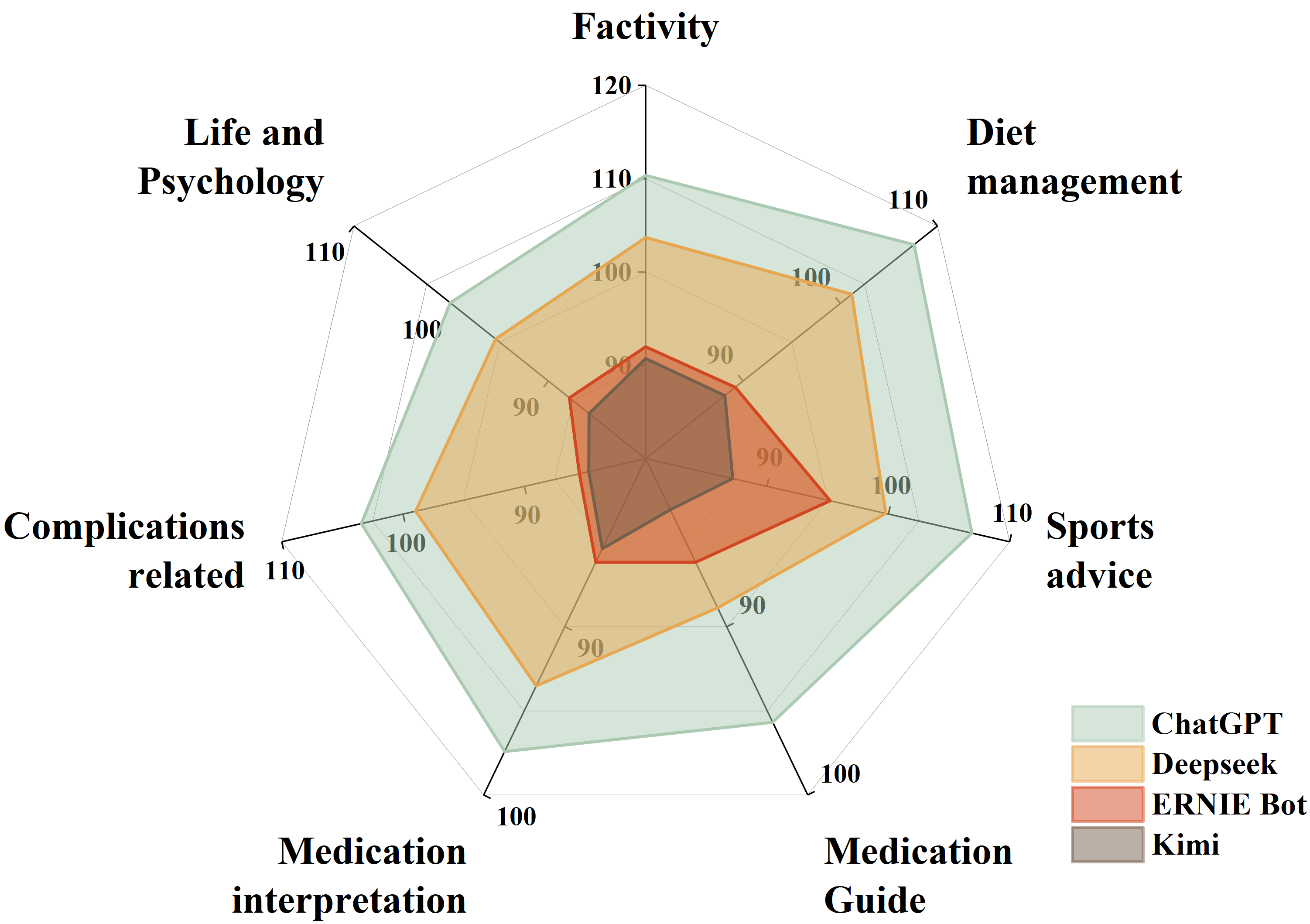}
    \caption{Overall comparison of AI performance across seven question categories.
    Each polygon represents one AI model, with larger enclosed areas indicating stronger performance across the corresponding category dimensions.}
    \label{fig:category_radar}
\end{figure}

Per-category analysis for each model (Figures~\ref{fig:category_performance} and~\ref{fig:category_radar}) \revise{showed} that ChatGPT maintained a consistently strong lead---particularly in \textit{\revise{Factual Knowledge}} ($M=110.52$), \textit{Diet Management} ($M=107.55$), \textit{Sports Advice} ($M=106.86$), and \textit{Complications Related} ($M=108.44$). DeepSeek performed comparably in factual topics (\textit{\revise{Factual Knowledge}} $M=108.63$) but dropped notably in \textit{Medication Interpretation} ($M=93.15$). Kimi's lowest rating appeared in \textit{Medication Guide} ($M=83.44$). ERNIE Bot showed relative strength in \textit{Life and Psychology} ($M=87.34$), where its more colloquial tone was perceived as accessible, though its factual reliability remained poor in safety-critical contexts.

\subsubsection{Model Strengths and Weaknesses by Category}

\begin{figure}[t]
\centering
\includegraphics[width=0.8\linewidth]{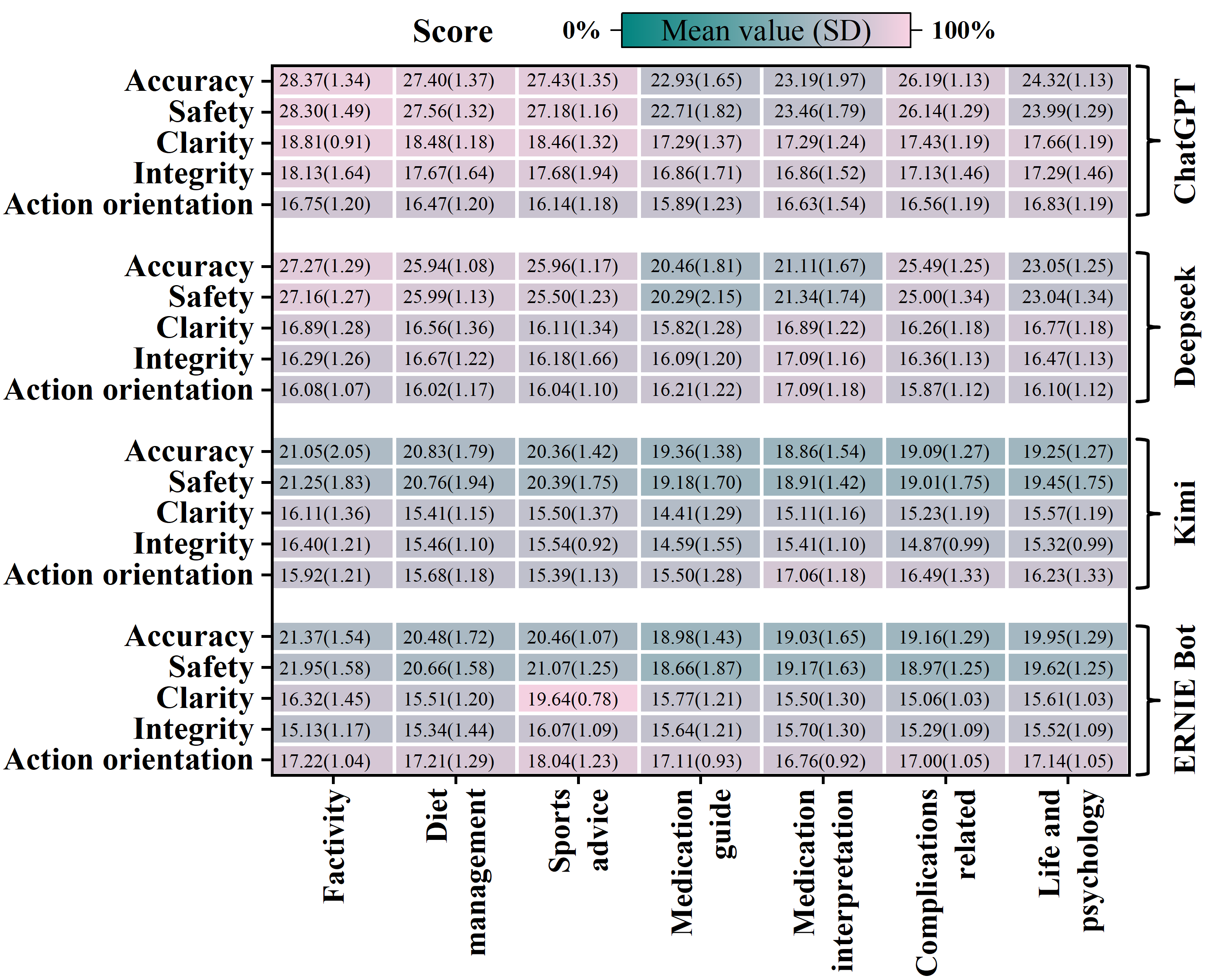}
\caption{Mean (SD) scores of each AI model across all combinations of evaluation dimensions and question categories. Color gradient encodes normalized score magnitude from 0\% to 100\%.}
\label{app:heatmap}
\end{figure}

The fine-grained heatmap (Figure~\ref{app:heatmap}) reveals asymmetries that explain model-specific behavior observed above. ChatGPT performed strongly across most combinations, but its \textit{Action Orientation} in \textit{Medication Interpretation} was notably low: while it could accurately describe insulin resistance or hypoglycemia mechanisms, it often stopped short of behavioral instructions such as adjusting monitoring frequency after a hypoglycemic episode.

DeepSeek exhibited solid \textit{Accuracy} and \textit{Clarity} for \textit{Diet Management} and \textit{\revise{Factual Knowledge}}, but its performance in \textit{Integrity} for \textit{Life and Psychology} was compromised by contradictory outputs---for example, simultaneously advising patients to ``monitor blood glucose at least four times daily for tight glycemic control'' while stating that ``frequent self-monitoring is unnecessary and may increase anxiety---once daily is sufficient for stable patients.'' Such within-response contradictions directly undermine physician trust.

Kimi displayed the most pronounced deficiencies in \textit{Clarity} and \textit{Integrity} across categories, often omitting key steps in procedural advice. ERNIE Bot's pattern was inverted: its consistently high \textit{Action Orientation}---strongest in \textit{Sports Advice} and \textit{Medication Guide}---occasionally lacked essential safety conditions, for instance omitting warnings for patients with retinopathy or neuropathy. This contrastive pattern---ERNIE Bot's directive clarity without factual safeguarding versus ChatGPT's accurate caution without behavioral specificity---highlights a persistent trade-off between actionability and safety that no single model currently resolves.

\subsection{Qualitative Findings}
\subsubsection{Perceptions of AI as a Health Information Source}

Physicians expressed cautious optimism about AI as a health information resource. They consistently recognized AI's strengths in factual explanation and patient education---particularly for newly diagnosed individuals. As P3 observed, \textit{``As an information aggregator, it performs remarkably well. I can imagine a newly diagnosed patient who knows nothing about `insulin resistance' coming in after consulting AI, asking more specific questions like, `Is my insulin resistance related to obesity?'''} This \textit{pre-visit primer} role was perceived as beneficial for raising baseline understanding and improving the efficiency of clinical encounters.

However, physicians drew a firm boundary between educational scaffolding and clinical authority. P2 stated, \textit{``I'm okay with patients bringing AI-printed materials to the clinic, but I can't accept AI directly telling a patient `your insulin dosage should be adjusted.'''} P4 elaborated: \textit{``AI sees a single data point; physicians see a multi-dimensional picture.''} This boundary became critical in queries approaching clinical decisions---as P5 noted, \textit{``It correctly listed the `dawn phenomenon' and `Somogyi effect' as causes of fasting hyperglycemia, but didn't suggest 3 a.m. blood glucose testing---which is key for differentiation.''} P6 identified a related safety gap: \textit{``When asked about ketogenic diets, AI gives pros and cons but fails to state clearly that it's contraindicated for patients with kidney disease. That kind of neutrality could mislead high-risk users.''}

Physicians also flagged omissions in otherwise thorough responses. P2 recalled: \textit{``It mentioned metformin side effects but downplayed that it must be stopped before contrast imaging. For elderly patients getting coronary CTs, this omission could be dangerous.''} They praised AI's communicative design---metaphors such as ``a key and a lock'' for insulin-receptor interaction (P7)---but noted that fluency did not always accompany completeness. As P3 put it, \textit{``It tells you `spinach is good' and `oatmeal is good,' but doesn't integrate that into a personalized plan that considers culture, budget, and comorbidities. It gives data points, not strategies.''}

Emotional responsiveness was a consistent failure point. P1 described overly scripted action guidance: \textit{``It says `test your blood sugar four times a day,' but won't say `test if you feel dizzy---even if it's unscheduled.'''} P6 was direct about the affective gap: \textit{``When a patient says `I'm terrified after diagnosis,' AI responds with checkbox lines like `stay positive, seek support.' Patients need to feel understood. AI can't do that.''}

\subsubsection{Comparative Evaluation of the Four AI Models}

Across all participants, ChatGPT was regarded as the most reliable and well-rounded system. P2 noted, \textit{``ChatGPT not only delivers accurate content but also includes physiological and clinical context---almost at the level of professional medical writing.''} P3 added that its safety-aware framing was specifically valued: \textit{``It tends to include appropriate disclaimers, like reminding users to consult a doctor. That kind of caution matters.''} Physicians attributed their greater confidence in ChatGPT to this combination of accuracy, structure, and calibrated tone.

DeepSeek was viewed as competent but clearly second-tier. It handled standardized factual queries well but struggled with ambiguity. P5 observed: \textit{``It performs well on textbook-style questions, but once contextual reasoning is needed, the gap with ChatGPT becomes clear.''} Notably, physicians found that DeepSeek could produce internally contradictory guidance in complex scenarios---particularly for medication and psychological topics---reducing its trustworthiness where precision matters most.

ERNIE Bot received mixed reviews. P7 characterized it as procedurally helpful but shallow: \textit{``It's good at listing steps---like how to rotate insulin injection sites---but the underlying explanation isn't always sound.''} Physicians saw its directive style as a double-edged sword: effective for simple behavioral prompts, but potentially misleading if users over-trust incorrect procedural advice.

Kimi was consistently rated as the weakest model across all five dimensions. P4 described a critical gap in an emergency context: \textit{``Its answer on hypoglycemia omitted basic steps---when to eat sugar, when to call for help. That kind of omission is unacceptable in health advice.''} Physicians uniformly indicated they would not deploy Kimi independently in any patient-facing scenario.

\subsubsection{Performance Patterns Across Question Types}

Physicians identified clear performance trends across question domains. Factual questions---biomedical definitions, diagnostic criteria, physiological mechanisms---played consistently to AI's strengths. P3 summarized: \textit{``For standardized questions like `What is HbA1c?' or `How does insulin resistance work?', AI---especially ChatGPT---delivers accurate, well-structured, accessible explanations. It's \revise{useful} as an entry-level educational tool.''} This educational framing was broadly endorsed for early-diagnosis contexts.

Lifestyle guidance revealed a gap between information breadth and actionable depth. P5 noted: \textit{``AI can list glycemic indices and general food principles, but when asked `How should a diabetic patient with kidney disease adjust their diet?', the answer is vague and lacks actionable specificity.''} Safety concerns also emerged: P6 emphasized that \textit{``AI reliably recommends `150 minutes of moderate exercise per week,' but often misses crucial warnings for patients with retinopathy or cardiovascular issues.''} Though broader AI caution was sometimes acknowledged---P4 remarked that its safety caveats were ``more thorough than some doctors''---this came with incomplete prioritization.

Medication-related questions marked a sharper decline in confidence. P1 drew the line clearly: \textit{``Explaining how a drug works is fine. But once the question becomes `Should I change my dose?' or `Can I switch medications?', AI quickly exceeds its capacity.''} P4 identified a structural problem: \textit{``AI often fails to prominently emphasize: `Any medication change must be discussed with a doctor.' That should be a non-negotiable message.''} Medication interpretation presented a related challenge: AI responses were often encyclopedic but provided no navigational guidance. P2 described this: \textit{``If a patient reports feeling constantly fatigued, AI lists everything from poor glucose control to thyroid disorders. But this list offers no direction---it just adds confusion.''}

Complications-related questions drew polarized reactions. Some physicians valued the cautious framing as awareness-raising; others found the language alarmist. P7 cautioned: \textit{``When describing complications like diabetic foot or retinopathy, the language often fails to convey that these are preventable and manageable---which may heighten patient anxiety rather than motivate engagement.''} In the psychological domain, AI was seen as superficially helpful but affectively hollow. P3 observed: \textit{``When it comes to things like diabetes burnout, the answers feel templated---patients can tell there's no real understanding behind them.''}

\subsubsection{Expectations for Future Health-AI Systems}

Physicians articulated four recurring expectations that together define what responsible AI health tools would need to provide. First, \textit{contextual personalization}: P6---\textit{``I want AI to stop answering like a textbook. If it could say `Because this patient has nephropathy, avoid high-protein diets,' that's when it becomes truly useful.''} Second, \textit{source transparency}: P2---\textit{``I need to know where the information is coming from. Was it a national guideline? A journal article? Without that, I can't calibrate how much weight to give it.''} Third, \textit{physiological data integration}---connecting CGM readings, step counts, or symptom logs to enable context-aware explanation. Fourth, explicit \textit{human-in-the-loop governance}: P7---\textit{``It should be a draft assistant. Let it write, but require a doctor to review before the patient sees anything involving medication changes.''} These expectations directly map onto the design directions developed in the Discussion.

\section{DISCUSSION}
\subsection{Generative AI as a Scalable Health Educator}

Our findings support the role of generative AI as an accessible tool for patient education in chronic disease management. Across both quantitative and qualitative analyses, physicians consistently acknowledged AI's ability to deliver structured, accurate explanations for factual knowledge and general lifestyle guidance, \revise{particularly for newly diagnosed patients}, aligning with prior work demonstrating LLM effectiveness in synthesizing biomedical knowledge \cite{akrimi2025chatgpt, hernandez2023future}.

\revise{This finding is grounded in participant reported information seeking.} Unlike prior evaluations based on static or expert-authored prompts \cite{shin2020autoprompt, mou2024sg}, our question corpus emerged from \revise{patient initiated information seeking} in naturalistic AI interactions. This grounding supports a more precise account of the function AI reliably performs. We use the term \textit{pre-visit primer} to name a preparatory role that is distinct from both formal patient education, which is \revise{clinician mediated} and typically occurs \revise{post diagnosis}, and shared \revise{decision making}, which requires co-present clinical judgment. \revise{We do not claim that pre visit information seeking is new. Our contribution is to show how GenAI reshapes this preparatory work in chronic care by offering immediate explanation while risking drift into clinical advice.} The pre-visit primer occupies a specific position in the patient \revise{information seeking} trajectory. It supports vocabulary acquisition and \revise{question formation} rather than definitive answers, and derives its value from being a step \textit{toward} clinical consultation rather than a substitute for it. Designing for this function implies different system priorities, \revise{including} surfacing appropriate uncertainty, scaffolding \revise{question formation}, and framing AI guidance as input for clinical discussion rather than as independent conclusions.

Our results also clarify the boundaries of this educational value. Even in its strongest domains, AI does not substitute for clinical reasoning. Physicians emphasized that factual explanations, no matter how clear, are not equivalent to judgment, \revise{a concern} consistent with prior critiques of how AI fluency may create the appearance of epistemic authority \cite{10.1145/3544548.3581251, rajagopal2025generativeaisupportpatients, asgari2025framework}.

\subsection{Performance Gaps Across Tasks and Models}

\revise{Despite the growing enthusiasm for deploying LLMs in health contexts, our results show performance differences between models and task categories. In this dataset,} ChatGPT \revise{received the highest physician ratings} in accuracy, clarity, and integrity. \revise{Kimi and ERNIE Bot exhibited frequent deficiencies}, especially in high-stakes tasks such as Medication Guide and Medication Interpretation. This aligns with prior work showing that LLMs vary widely in factual correctness and linguistic reliability \cite{samimi2025visual, hernandez2023future,su2026capnav,zhang2025slideaudit}, while extending these findings through physician-rated evaluation across clinically relevant question types rather than a single model in isolation \cite{10.1145/3706598.3713819, gaber2025evaluating, jahan2024comprehensive}.

Our task-level analysis revealed a domain-dependent performance gradient. Factual and lifestyle-related questions were answered reliably by most models, while categories requiring clinical reasoning, \revise{including medication adjustment and symptom interpretation}, or emotional sensitivity showed consistent underperformance, \revise{consistent with} qualitative evidence from prior studies on LLM limitations in structured clinical reasoning \cite{antonie2025role, singhal2025toward, yang2024kg}. Physicians uniformly cautioned against deploying these tools in high-risk scenarios, pointing specifically to omissions such as missing dosage warnings or failing to flag contraindications. \revise{These examples show how} surface-level fluency may mask substantive clinical gaps \cite{10.1145/3706598.3713497, reis2024influence, abdelwanis2024exploring}. Specific design responses to these gaps are discussed in Section~6.4.

\subsection{Designing for Trust and Appropriate Use Boundaries}

Our findings highlight a distinction between appropriate trust and overreliance in the use of generative AI for health information. Physicians recognized AI's educational value but uniformly rejected it as a clinical authority, \revise{which is} consistent with broader concerns in human-centered AI research that system usability and fluency may inadvertently foster misplaced confidence \cite{rosbach2025automation, shekar2024people, wysocki2023assessing,meng2026living}. Prior research has explored how interface transparency, uncertainty expression, and provenance cues can support trust calibration in AI-generated content \cite{reyes2025trusting, bhatt2021uncertainty}. \revise{Our physician participants made similar demands, pointing specifically to the absence of source traceability and inadequate safety framing in current AI responses.}

A more specific mechanism emerged from our qualitative data. As P4 explained, \textit{``It might sound professional, but sometimes it misses the one thing that truly matters---the context that makes an answer safe.''} \revise{We use \textit{fluency illusion} as a health specific lens on established concerns about automation bias, overtrust, hallucination, and perceived authority.} \revise{We define this pattern as} the tendency for well structured, confident language to create an impression of epistemic authority that the underlying content may not warrant. This \textit{fluency illusion} poses particular challenges in consumer health contexts. It parallels concerns about automation bias in human-AI interaction \cite{rosbach2025automation, shekar2024people}, but operates with greater force when users lack the domain expertise to detect errors independently. Unlike professionals in other high-stakes domains who may catch system failures through domain knowledge, lay patients interpreting health information have limited means to distinguish linguistic fluency from clinical validity. The underlying cognitive pathway runs from perceived linguistic quality to inferred epistemic authority to uncritical action. \revise{In this sequence,} rhetorical surface substitutes for independent evaluation of clinical content. \revise{This framing narrows the mechanism rather than presenting the phenomenon as wholly new.} Whereas automation bias describes a tendency to defer to system outputs over human judgment \cite{rosbach2025automation, shekar2024people}, the fluency illusion operates within the output itself. \revise{Users are not simply trusting a system over a person. They are unable to detect that well formed language does not entail well grounded claims.} Designing against the fluency illusion therefore goes beyond improving accuracy to include interface-level mechanisms that make the distinction between language quality and epistemic reliability legible to non-expert users.

\revise{These observations point to specific design challenges. Interfaces should} frame AI outputs as provisional rather than prescriptive, distinguish educational summaries from clinical recommendations, and trigger mandatory safety disclaimers or referral prompts for high-stakes topics. Building on prior work on uncertainty communication and explainability \cite{merry2021mental, schrills2020answer, bauer2023expl,luo2025s}, such mechanisms may help cultivate calibrated trust, where users neither dismiss AI advice outright nor accept it uncritically, without reducing AI's practical utility.

\subsection{Limitations in AI's Capacity for Human-Centered Information Interfaces}

While generative AI demonstrates strong potential in factual information delivery, our findings point to important limitations in its capacity as a health information interface. Personalization was a core failure. \revise{Physicians observed} that while models could accurately define terms or enumerate dietary principles, responses consistently failed to adapt to individual comorbidities or behavioral constraints. Generic guidance such as ``eat a balanced diet'' was perceived as inadequate or misleading when key risk factors went unacknowledged, \revise{a finding} that resonates with long-standing calls in HCI and health informatics for user-adaptive, contextually grounded health technologies \cite{lewis2022can, torkamaan2022recommendations, jacobs2018mypath,meng2026engagement}.

Actionability was a second persistent limitation. Physicians found that AI systems frequently listed options rather than offering navigable guidance, \revise{such as enumerating} possible causes of fatigue without indicating what to do next, or describing drug mechanisms without translating them into behavioral steps \cite{he2025conversational, song2025interaction,liu2025supporting}. Emotional responsiveness was a third failure point. \revise{Responses} to affect-laden prompts were consistently described as formulaic, \revise{polished in language but perceptibly hollow} in emotional attunement \cite{pan2025developing, sun2025interface, xiao2023powering,zhao2025immersive}. Our analysis suggests that current generative AI tools, while linguistically competent, function primarily as information delivery systems rather than adaptive interlocutors, \revise{a limitation} that is particularly consequential in chronic care contexts where patients require not just accurate information but sustained, context-sensitive engagement.

\subsection{Converging and Diverging Perspectives Across Patients and Physicians}

\revise{This comparison is not a same stimulus evaluation. Patients did not rate the exact AI outputs evaluated by physicians. Instead, we compare patient reported experiences and needs from Study~1 with physician evaluations of AI responses to a curated subset of patient derived questions in Study~2.}

A defining feature of our mixed-methods design is that patients and physicians evaluated AI-generated health information independently. This independence makes convergences particularly meaningful: when both groups reach the same conclusion through different evaluative criteria, the conclusion carries empirical robustness that neither perspective alone could provide.

Three convergences stand out. First, both groups independently drew the same boundary around AI's role. Patients consistently deferred to physician authority when AI conflicted with clinical advice (\textit{``If AI says one thing and my doctor says another, I will always follow the doctor''}), positioning AI as an orientational resource rather than a decision-making agent. Physicians arrived at the same boundary from the opposite direction, welcoming AI's educational function while uniformly rejecting it as a clinical authority (P2: \textit{``I'm okay with patients bringing AI-printed materials to the clinic, but I can't accept AI directly telling a patient their insulin dosage should be adjusted''}). Neither group was primed toward this position. \revise{Both reached it independently, suggesting} that the educator-versus-decision-maker distinction is not a researcher's theoretical construct but a boundary that both patient and clinical stakeholders recognize in practice.

Second, both groups independently identified emotional support as AI's most critically underdeveloped capability. In Study~1, psychological support received patients' lowest attitude ratings ($M=5.10$, $SD=1.41$), \revise{a full four points below convenience scores}. In Study~2, physicians described AI's emotional outputs in near-identical terms, \revise{as formulaic and perceptibly hollow} (P6: \textit{``When a patient says `I'm terrified after diagnosis,' AI responds with checkbox lines like `stay positive, seek support.' But patients need to feel understood''} \revise{and} P4: \textit{``It might say `I understand your anxiety,' but there's no feeling behind those words. Patients can tell''}). The fact that patients registered this deficit in self-reported ratings while physicians independently articulated the same critique through clinical reasoning suggests emotional inadequacy is structurally visible, \revise{and legible} to lay users without clinical training.

Third, personalization emerged as a shared unmet need. Patients rated AI's personalization capacity at $M=5.62$, with many requesting systems that account for individual comorbidities and lived constraints (P3, Study~1: \textit{``I hope AI can truly understand my condition, not just repeat a fixed template''}). Physicians independently identified the same gap from a clinical standpoint (P4: \textit{``AI sees a single data point; physicians see a multi-dimensional picture''} \revise{and} P6: \textit{``It doesn't integrate advice into a personalized plan that considers culture, budget, and comorbidities''}).

Two meaningful divergences also emerged. First, patients and physicians differed in their primary evaluative concern. Patients' most positive ratings clustered around convenience and accessibility ($M=9.05$ and $M=8.19$, respectively), reflecting that for patients, AI's core value proposition is availability, \revise{an always accessible resource} that lowers the threshold for health engagement. Physicians were primarily attuned to safety gaps, \revise{including} omitted warnings, misleading framing, and risks of patients acting on incomplete advice. This divergence reflects different baseline positions. \revise{Patients evaluate AI against having no information, whereas physicians evaluate it against the standard of clinical adequacy.} Effective design must address both simultaneously.

Second, the two groups calibrated trust through fundamentally different mechanisms. Patients tended to trust AI based on perceived helpfulness and linguistic responsiveness. Physicians evaluated trust through clinical accuracy, safety signaling, and the presence of appropriate disclaimers. The same AI response that satisfies a patient's informational need may simultaneously alarm a physician who recognizes what is missing, \revise{an asymmetry} that underscores the need for layered interfaces that present accessible content to patients while making clinical caveats more salient for high-risk information.

\subsection{Design Implications for AI-Mediated Health Support in Chronic Care}

Building on findings across both studies, we outline four design directions for AI-mediated health tools in chronic care, \revise{each linked to specific performance gaps in our analysis}.

First, \textit{task-aware model orchestration}. As no single model performs consistently across all content types and risk levels, AI health tools should support query routing based on task demands \cite{cao2026causalinfluencemaximizationsteadystate}. \revise{A practical interface could first classify a query as factual education, lifestyle guidance, medication decision, symptom interpretation, complication concern, or emotional support. Low risk factual queries could enter an educational mode. Medication and symptom interpretation queries should enter a safety constrained mode with source display, clarification prompts, and escalation options.} Our finding that performance drops approximately ten points from factual categories ($M=99.19$) to medication-related categories ($M=88.44$) reflects fundamentally different reasoning requirements that call for different system behaviors \cite{naik2025designing, lee2025map,ma2026can,liu2025supporting}. Notably, ERNIE Bot's Action Orientation ($M=17.14$), \revise{the highest among all four models and above ChatGPT's score} ($M=16.52$), illustrates that directive, procedural output styles carry task-specific value. \revise{Such strengths could be leveraged in procedural self care scenarios under appropriate safety constraints}, rather than treating any single model as universally optimal.

Second, \textit{risk-aware fallback strategies}. For sensitive domains such as symptom triage or medication decisions, interfaces should detect uncertainty and activate protective flows, \revise{including high visibility disclaimers}, clarifying questions, or escalation pathways to human professionals \cite{florins2004graceful, bhatt2021uncertainty, reyes2025trusting,meng2026balancing}. \revise{Fallback should be triggered by dose change verbs, drug names, insulin adjustment, severe or ambiguous symptoms, hypoglycemia or hyperglycemia episodes, kidney or cardiovascular comorbidity, and complications such as retinopathy or neuropathy. A safe fallback response should avoid individualized treatment instructions, explain why the request depends on clinical context, provide general educational information, and direct the patient to clinical care when risk is high.} This direction responds directly to physicians' most consistent concern. \revise{Even} ChatGPT showed its largest Action Orientation deficit in Medication Interpretation, \revise{the domain where} the absence of actionable guidance carries the highest patient risk, and where physicians uniformly noted that safety framing was either absent or insufficiently prominent.

Third, \textit{dynamic personalization}. Systems should move beyond static profiles to incorporate \revise{real time} behavioral and physiological signals, \revise{including} wearable data, symptom logs, and medication history, to adapt recommendations to individual context \cite{lee2025map, ayobi2021co,meng2026living,meng2026engagement}. \revise{Personalization should be consent based and bounded. Systems should make clear which user context is known, unknown, or too clinically consequential for automated advice.} This direction is supported by convergent evidence from both studies. \revise{Patients' lowest personalization ratings} ($M=5.62$) and physicians' strongest qualitative critiques (P4: \textit{``AI sees a single data point; physicians see a multi-dimensional picture''}) identify the same deficit independently, making contextual adaptation the most robustly evidenced design target in our study.

Fourth, \textit{emotionally attuned interaction}. AI health tools should better attend to the emotional register of patient queries \revise{through affect aware response patterns}, structured validation, or clear pathways to human support \cite{gencc2024situating, cuadra2024illusion,meng2026creating,meng2026focus}. \revise{For affect laden prompts, systems should combine validation, normalization, practical coping steps, and routes to human support. The goal is not to simulate clinical empathy as a substitute for care, but to avoid dismissive template responses.} Patients' lowest-rated attitude dimension ($M=5.10$ for psychological support) and physicians' consistent characterization of AI emotional responses as formulaic (P6: \textit{``checkbox lines''}, P4: \textit{``patients can tell''}) converge on this gap from independent directions, making emotional responsiveness the dimension where both patient experience data and clinical judgment most directly indicate the need for design improvement.

\subsection{Limitations and Future Work}

This study has several limitations. First, AI model performance changes rapidly. \revise{The outputs we analyzed represent a snapshot}, and subsequent fine-tuning or safety alignment may alter performance characteristics. Longitudinal tracking of AI evolution \revise{would help} monitor emerging strengths and risks in deployment. Second, our evaluation was limited to \revise{static, single turn outputs}. \revise{Real world} health interactions are dynamic, often involving follow-up clarification, user error, and correction. \revise{Future work should extend this framework to simulate multi-turn interactive scenarios that better reflect the complexity of AI-mediated health communication.} \revise{Future work should also ask patients and clinicians to evaluate the same AI responses to test whether the convergences observed here persist under a same stimulus design.} Third, the physician sample comprised seven endocrinologists affiliated with Chinese hospitals, limiting generalizability across provider types, clinical contexts, and healthcare systems. Broader sampling, \revise{including primary care practitioners, nurses, and rural providers}, could reveal different expectations regarding AI trustworthiness and workflow integration. \revise{The query corpus was participant reported and partly based on paraphrased recollections, so it should not be read as a complete interaction log. The five dimensional rubric was a structured expert evaluation tool rather than a psychometrically validated scale.} Finally, we did not employ a formalized prompt engineering protocol. \revise{Given the known sensitivity of LLMs to input phrasing, future research should systematically examine how prompt variation affects both performance and user experience in health-critical domains.}

\section{CONCLUSION}
This paper presents a mixed-methods investigation into the quality and limitations of AI-generated health information for T2DM self-management, combining an observational study of 21 patients with a structured evaluation by seven physicians across four commercial AI systems. The study \revise{identifies key tensions in how systems balance} accessibility with clinical appropriateness, standardization with personalization, and linguistic fluency with factual validity, \revise{which characterize} the current state of generative AI in chronic disease contexts.

Our findings \revise{show} consistent AI strengths in factual explanation and general lifestyle guidance, alongside persistent deficiencies in medication reasoning, contextual personalization, and emotional responsiveness. The five-dimensional evaluation \revise{rubric}, assessing \textit{Accuracy, Safety, Clarity, Integrity}, and \textit{Action Orientation}, \revise{differentiated physician ratings across models and question categories in this dataset}, and may offer a \revise{useful structure} for AI health information evaluation in related chronic care contexts.

Two \revise{analytic concepts} emerge from the data. The \textit{pre-visit primer} frames AI's most defensible role as preparatory rather than substitutive, helping patients orient toward clinical encounters rather than bypass them. The \textit{fluency illusion} identifies how linguistically polished AI outputs may convey epistemic authority that the underlying clinical content does not support. Across both studies, patients and physicians independently converged on the same three limitations, role boundaries, emotional inadequacy, and personalization gaps, while diverging in their primary evaluative concerns, together providing \revise{triangulated evidence} for the four design directions this paper develops.

Chronic conditions demand more of AI than accurate recall. \revise{They require} context-sensitive, emotionally attuned support that operates \revise{within existing clinical relationships and in deference to them}. This work offers empirically grounded design directions and \revise{two analytic concepts} for that longer-term challenge.

\section{GenAI Usage Disclosure}
In preparing this manuscript, we used ChatGPT in limited ways to support formatting and consistency. Specifically, the tool was employed to assist in generating Overleaf code for demographic tables and detailed tables in the Appendix, to help transform quantitative and qualitative interview materials into more readable Appendix formats, and to check grammar and terminology consistency across sections. The use of AI was restricted to these supportive tasks only. All authors take full responsibility for the content, analysis, and claims made in this paper. All data are real, and all perspectives and interpretations presented are those of the authors; no fabricated data or non-genuine content were produced through the use of large language models.

\bibliographystyle{ACM-Reference-Format}
\bibliography{Aref}

\section{Appendix}
\appendix

\section{Attitudes Toward AI Questionnaire}

Participants rated the following items on a 10-point Likert scale, 
where 1 indicates \textit{Strongly Disagree} and 10 indicates \textit{Strongly Agree}.

\renewcommand{\arraystretch}{1.3}
\begin{longtable}{c p{0.46\linewidth} c c c c c}
\caption{Attitudes Toward AI Questionnaire (10-point Likert scale)} \label{tab:ai_attitudes} \\
\toprule
\textbf{ID} & \textbf{Statement} & \textbf{\small SD} & \textbf{\small D} & \textbf{\small N} & \textbf{\small A} & \textbf{\small SA} \\
\midrule
\endfirsthead
\caption[]{Attitudes Toward AI Questionnaire (continued)} \\
\toprule
\textbf{ID} & \textbf{Statement} & \textbf{\small SD} & \textbf{\small D} & \textbf{\small N} & \textbf{\small A} & \textbf{\small SA} \\
\midrule
\endhead
\bottomrule
\multicolumn{7}{r}{\small SD=Strongly Disagree \quad D=Disagree \quad N=Neutral \quad A=Agree \quad SA=Strongly Agree}
\endfoot
Q1 & AI provides information in a very convenient and timely manner, allowing me to obtain answers anytime and anywhere. & & & & & \\[4pt]
Q2 & I believe the information provided by AI is accurate and reliable. & & & & & \\[4pt]
Q3 & AI’s responses are expressed in plain language, with clear structure, making them easy for me to understand and follow. & & & & & \\[4pt]
Q4 & AI can provide personalized advice tailored to my specific condition (e.g., disease course, symptoms), rather than only general answers. & & & & & \\[4pt]
Q5 & After consulting AI, I feel more confident in managing my condition. & & & & & \\[4pt]
Q6 & AI’s responses help me organize my thoughts, enabling more effective communication with my doctor. & & & & & \\[4pt]
Q7 & I worry that over-reliance on AI may cause me to ignore important bodily signals or delay seeking medical care. & & & & & \\[4pt]
Q8 & Interacting with AI provides me with a certain degree of psychological comfort and support when I feel anxious or confused. & & & & & \\[4pt]
Q9 & AI can objectively compare different treatment options (e.g., medication, lifestyle adjustments) and highlight their advantages and disadvantages. & & & & & \\[4pt]
Q10 & Overall, I am satisfied with using AI as an auxiliary tool for diabetes management. & & & & & \\
\bottomrule
\end{longtable}
\renewcommand{\arraystretch}{1}

\section{AI Usage Questionnaire}

This questionnaire was designed to systematically capture the actual use of AI platforms by patients with T2DM.  
Platforms are categorized by functionality (general-purpose vs. healthcare-specific) and further distinguished by access modality (standalone apps, system-integrated AI, wearable/embodied devices, etc.).  
Participants were asked to indicate whether they had used each platform.

\renewcommand{\arraystretch}{1.2}
\begin{longtable}{l l p{0.28\linewidth} p{0.28\linewidth}}
\caption{AI Usage Questionnaire} \label{tab:ai_usage} \\
\hline
\textbf{Function} & \textbf{Type} & \textbf{Platform} & \textbf{Access Modality} \\
\hline
\endfirsthead
\caption[]{AI Usage Questionnaire (continued)} \\
\hline
\textbf{Function} & \textbf{Type} & \textbf{Platform} & \textbf{Access Modality} \\
\hline
\endhead
\hline
\endfoot
\multicolumn{4}{l}{\textbf{General-purpose AI}} \\
& Large Language Models & ChatGPT & Website \\
& & DeepSeek & App, Website \\
& & Wenxin Yiyan & App, Website \\
& & Doubao & App, Website \\
& & Tongyi Qianwen & App, DingTalk-integrated, Website \\
& & Zhipu Qingyan & App, Website \\
& & Kimi Chat & App, Website \\
& & iFlytek Spark & App, Website \\
\multicolumn{4}{l}{\textbf{System-integrated AI}} \\
& & Tencent Yuanbao & App (Mini Program) \\
& & SenseChat & App, Website \\
& & Xiao Ai Assistant & Xiaomi phone / smart speaker \\
& & Vivo Blue Heart Assistant & vivo / iQOO phones \\
& & OPPO Breeno AI Assistant & OPPO / OnePlus / Realme phones \\
& & Huawei Xiaoyi & Huawei phones \\
\multicolumn{4}{l}{\textbf{Other General-purpose}} \\
& & Tiangong AI & App, Website \\
& & 360 Zhinao & App, Website \\
\multicolumn{4}{l}{\textbf{Healthcare AI}} \\
& General-purpose Medical AI & DingXiang Doctor & App, Mini Program \\
& & Chunyu Doctor & App, Mini Program \\
& & HaoDF Online & App, Mini Program \\
& & Ping An Health & App, Mini Program \\
& & WeDoctor & App, Mini Program \\
& Diabetes-specific AI & Sugar Nurse & App \\
& & Weitang & App \\
& & Glucose Manager & App \\
\multicolumn{4}{l}{\textbf{Embodied Interfaces}} \\
& & Tmall Genie & Smart speaker \\
& & Apple Watch & Smartwatch / wristband \\
& & Huawei Watch & Smartwatch / wristband \\
& & Xiaomi Band & Smartwatch / wristband \\
& & In-car AI & Vehicle interface \\
\multicolumn{4}{l}{\textbf{Other}} \\
& & \multicolumn{2}{l}{Please specify any additional AI platform related to T2DM that you use.} \\
\hline
\end{longtable}
\renewcommand{\arraystretch}{1}

\newcolumntype{L}[1]{>{\RaggedRight\arraybackslash}p{#1}}

\section{Semi-structured Interview Guideline}

The semi-structured interview guideline was designed to complement the questionnaires and textual records, ensuring rich and contextual insights. 
Each dimension included a guiding question and follow-up probes to encourage participants to elaborate.

\begin{longtable}{L{3.8cm} L{6.2cm} L{6.2cm}}
\caption{Semi-structured Interview Guideline for Patients with T2DM} \\
\toprule
\textbf{Interview Dimension} & \textbf{Guiding Question} & \textbf{Follow-up Probes / Key Focus} \\
\midrule
\endfirsthead

\caption[]{Semi-structured Interview Guideline for Patients with T2DM (continued)} \\
\toprule
\textbf{Interview Dimension} & \textbf{Guiding Question} & \textbf{Follow-up Probes / Key Focus} \\
\midrule
\endhead

\bottomrule
\endfoot

\textbf{1. Patient Background and Disease Management Habits} &
Can you describe your personal background and your current approach to managing T2DM? &
\par\textbullet\ Demographic information (age range, gender, urban/rural residence, education level, self-assessed ability to search health information online).%
\par\textbullet\ Disease characteristics (years since diagnosis, self-evaluation of current glycemic control such as HbA1c, diagnosed complications, and their impact on daily life).%
\par\textbullet\ Current management model (treatment type such as oral medication or insulin, blood glucose monitoring methods and frequency, adherence to diet and exercise management). \\

\textbf{2. AI Usage Behaviors and Question Content} &
How do you use AI platforms when managing your diabetes? &
\par\textbullet\ Breadth and depth of AI use (use of generative AI such as Wenxin Yiyan, ChatGPT, or voice assistants such as Siri or Xiao Ai; frequency and duration of use; scenarios such as symptom onset, before follow-up visits, or daily learning).%
\par\textbullet\ Platform choice and switching behaviors (most frequently used AI platform, reasons for selection such as reputation, usability, cost, or answer quality; whether multiple platforms were compared).%
\par\textbullet\ Specific question types (diet management: glycemic index, meal planning, dining out; medication management: side effects, dosage, missed dose handling, interactions, insurance; exercise prescriptions: type, intensity, contraindications; symptom interpretation: fluctuations, abnormal values, lab results; complications: prevention and treatment; lifestyle and psychology: stress management, communication with family). \\

\textbf{3. Attitudes Toward AI in Healthcare Information Provision} &
What is your perception of AI’s role in providing healthcare information? &
\par\textbullet\ Perceived information quality (accuracy, comprehensiveness, timeliness, clarity, consistency with physician advice).%
\par\textbullet\ Service experience (convenience, response speed, personalization, naturalness of interaction).%
\par\textbullet\ Utility and impact (improvement in knowledge, self-management confidence, and communication efficiency with physicians).%
\par\textbullet\ Risk perception and trust (concerns about misinformation, privacy and data security, and lack of empathy).%
\par\textbullet\ Role positioning (AI as supplement vs. replacement vs. competitor to doctors; types of issues requiring physician consultation; expected supportive roles such as pre-consultation data collection or post-consultation follow-up). \\

\textbf{4. Deeper Insights and Future Expectations} &
How do you envision future AI health assistants? &
\par\textbullet\ Response to conflicts between AI advice and physician recommendations.%
\par\textbullet\ Desired features: integration with personal health devices (glucometers, fitness trackers), image recognition (food, wounds), provision of guideline-based information with clear references, expert review and certification of AI responses, improved emotional and psychological support capabilities. \\

\bottomrule
\end{longtable}

\section{List of Participant Questions}

\begin{longtable}{p{2.8cm}p{1.8cm}p{8.2cm}p{1.5cm}}
\caption{List of Participant Questions (Appendix 4).} \\
\toprule
\textbf{Category} & \textbf{Question ID} & \textbf{Question Content} & \textbf{Frequency} \\
\midrule
\endfirsthead

\caption[]{List of Participant Questions (continued).} \\
\toprule
\textbf{Category} & \textbf{Question ID} & \textbf{Question Content} & \textbf{Frequency} \\
\midrule
\endhead

\bottomrule
\endfoot

\bottomrule
\endlastfoot

\textbf{\revise{Factual Knowledge}} & 1 & How did I get T2DM? & 17 \\
 & 2 & Is T2DM caused by eating too much sugar? & 10 \\
 & 3 & Can T2DM be cured? & 19 \\
 & 4 & What is insulin resistance? & 7 \\
 & 5 & Why does my body develop resistance? & 7 \\
 & 6 & Will this disease be inherited by my children? & 8 \\
 & 7 & What does prediabetes mean? Should it be treated? & 6 \\
 & 8 & What are carbohydrates, and how are they related to sugar? & 6 \\
 & 9 & What is HbA1c, and why is this test necessary? & 11 \\

\textbf{Diet Management} & 1 & Is the diabetic diet special? & 13 \\
 & 2 & What foods can’t I eat? Should I avoid all sugar and staples? & 17 \\
 & 3 & What foods can lower blood sugar? Are there “superfoods”? & 18 \\
 & 4 & Can I eat fruit? Which kinds are suitable, and when should I eat them? & 13 \\
 & 5 & How many meals a day should I have? How should I plan portions? & 18 \\
 & 6 & Can I eat sugar-free or substitute foods freely? & 12 \\
 & 7 & Are artificial sweeteners safe? & 5 \\
 & 8 & Is eating small frequent meals better for blood sugar control? & 5 \\
 & 9 & How should I choose food when dining out? & 8 \\
 & 10 & What foods look healthy but actually contain hidden sugar? & 7 \\
 & 11 & Why does porridge cause a rapid rise in blood sugar? & 5 \\
 & 12 & How can I calculate carbohydrates in foods? Are there easy methods? & 12 \\
 & 13 & What are glycemic index and glycemic load? How can I use them? & 11 \\
 & 14 & How do protein and fat affect blood sugar? Can I eat more of them? & 11 \\

\textbf{Sports Advice} & 1 & What kind of exercise suits me best, and how long should I do it? & 17 \\
 & 2 & When is the best time to exercise—before or after meals? & 5 \\
 & 3 & What should I do if my blood sugar drops during exercise? & 2 \\
 & 4 & What exercise restrictions exist when complications are present? & 2 \\

\textbf{Medication Guide} & 1 & Do I have to take medicine, or can I manage through diet and exercise alone? & 13 \\
 & 2 & Does starting insulin mean my condition is severe or that I’ll become dependent? & 10 \\
 & 3 & After diagnosis, do I need lifelong medication? & 15 \\
 & 4 & What are the side effects of oral drugs and insulin? & 3 \\
 & 5 & Why must I still control my diet while on medication? & 12 \\
 & 6 & How should I rotate injection sites properly? & 4 \\
 & 7 & What types of insulin exist, and what are their differences? & 3 \\
 & 8 & Under what circumstances should insulin therapy be initiated? & 3 \\

\textbf{Medication Interpretation} & 1 & Why is my blood sugar unstable even with regular treatment? & 9 \\
 & 2 & What is the recommended range for blood sugar levels? & 16 \\
 & 3 & Do I need to test blood sugar every day? When is the best time? & 16 \\
 & 4 & My blood sugar is slightly high but I feel fine—do I need treatment? & 11 \\
 & 5 & Why does my blood sugar rise after exercise? & 11 \\
 & 6 & How often should I test blood sugar, and at what time is it most accurate? & 12 \\
 & 7 & What readings are considered normal on a glucometer? & 13 \\
 & 8 & Why is my fasting blood sugar always high? & 12 \\
 & 9 & Why do I always feel tired or weak? & 11 \\
 & 10 & Why is my post-meal blood sugar harder to control than fasting levels? & 4 \\

\textbf{Complications Related} & 1 & Why does diabetes increase cardiovascular risk? & 2 \\
 & 2 & What are the main complications of diabetes, and how severe are they? & 12 \\
 & 3 & How can I distinguish diabetic ketoacidosis from hyperosmolar hyperglycemia? & 12 \\
 & 4 & How often should I have my eyes examined for retinopathy? & 5 \\
 & 5 & What is diabetic foot, and how can I prevent it? & 5 \\
 & 6 & Do hypertension or lipid-lowering drugs conflict with diabetes medications? & 5 \\
 & 7 & Is glucose variability more harmful than persistent hyperglycemia? & 5 \\
 & 8 & Will I inevitably develop complications? How can I prevent them? & 11 \\
 & 9 & What is the relationship between T2DM and cardiovascular diseases? & 7 \\
 & 10 & Why does T2DM affect sexual function? & 5 \\

\textbf{Life and Psychology} & 1 & How should I manage my blood sugar when I’m sick (e.g., cold, fever)? & 11 \\
 & 2 & How can I handle discrimination or stigma related to diabetes? & 10 \\
 & 3 & How can I carry and store insulin while traveling or on business trips? & 12 \\
 & 4 & Can people with diabetes drive, and what precautions should they take? & 7 \\
 & 5 & Does poor sleep affect blood sugar levels? & 7 \\
 & 6 & How can I cope with fear and anxiety after diagnosis? & 6 \\
 & 7 & What is diabetes burnout, and how can I manage it? & 12 \\
 & 8 & How should I communicate my condition with family members? & 7 \\
 & 9 & How can I set realistic long-term health goals? & 7 \\
 & 10 & How often should I have checkups, and which tests are necessary? & 17 \\
 & 11 & What are the effects of smoking and drinking on blood sugar? & 2 \\

\end{longtable}

\newcolumntype{L}[1]{>{\RaggedRight\arraybackslash}p{#1}}

\section{Evaluation Interview Framework}

This framework summarizes the semi-structured evaluation interviews conducted with physicians in the evaluative study. 
The goal was to assess physicians’ perceptions, evaluation criteria, and expectations toward AI-generated information for T2DM management. 
Each dimension included a guiding question and follow-up probes to encourage elaboration.

\begin{longtable}{L{3.8cm} L{6.2cm} L{6.2cm}}
\caption{Semi-structured Evaluation Interview Framework with Physicians} \\
\toprule
\textbf{Interview Dimension} & \textbf{Guiding Question} & \textbf{Follow-up Probes / Key Focus} \\
\midrule
\endfirsthead

\caption[]{Semi-structured Evaluation Interview Framework with Physicians (continued)} \\
\toprule
\textbf{Interview Dimension} & \textbf{Guiding Question} & \textbf{Follow-up Probes / Key Focus} \\
\midrule
\endhead

\bottomrule
\endfoot

\textbf{1. Overall Impression and Trust} &
How would you describe your overall impression of the AI-generated T2DM information?  
Under what circumstances would you use or avoid it? &
\par\textbullet\ Initial trust boundaries and willingness to rely on AI outputs.%
\par\textbullet\ Perceived strengths and weaknesses of AI-generated content. \\

\textbf{2. Accuracy and Safety Assessment} &
How accurate and safe is the information compared to your clinical knowledge or official guidelines? &
\par\textbullet\ Examples of factual errors, outdated or misleading statements.%
\par\textbullet\ Whether risk warnings are sufficient and appropriately phrased.%
\par\textbullet\ Potential for patient misunderstanding or harmful interpretation. \\

\textbf{3. Clarity, Completeness, and Actionability} &
How clear, complete, and actionable are the AI responses from a clinical and patient-education perspective? &
\par\textbullet\ Clarity for patients with different education levels.%
\par\textbullet\ Completeness of topic coverage (e.g., diet, exercise, complications).%
\par\textbullet\ Specificity and feasibility of behavioral recommendations. \\

\textbf{4. Evaluation by Content Category} &
How would you assess the quality and appropriateness of AI responses across different question types? &
\par\textbullet\ Factual knowledge: accuracy of core medical concepts (e.g., insulin resistance, HbA1c).%
\par\textbullet\ Dietary advice: safety, personalization, and clarity (e.g., sugar-free foods, carb counting).%
\par\textbullet\ Sports Advice: suitability and safety for comorbid conditions.%
\par\textbullet\ Medication Guide: adequacy of warnings about adherence and risks.%
\par\textbullet\ Medication Interpretation: precision and contextual relevance.%
\par\textbullet\ Complications Related: balance between warning and reassurance.%
\par\textbullet\ Life and Psychology: empathy and practicality of coping suggestions. \\

\textbf{5. Practical Application Scenarios} &
In which professional or patient-education contexts could this AI-generated content be safely and effectively used? &
\par\textbullet\ Appropriate use cases (e.g., pre-consultation education, clinician reference, health outreach).%
\par\textbullet\ “Red-line” scenarios where AI content should not be used (e.g., replacing diagnosis, altering treatment).%
\par\textbullet\ Additional disclaimers or contextual notes needed before showing to patients. \\

\textbf{6. Comparative Evaluation and Improvement} &
Compared with human-written materials, how would you evaluate AI’s advantages and shortcomings?  
What improvements would you prioritize? &
\par\textbullet\ Comparative efficiency, cost, and flexibility of AI vs. human-created content.%
\par\textbullet\ Priority areas for improvement (accuracy, personalization, empathy, safety).%
\par\textbullet\ Willingness to participate in future evaluations and collaboration modes. \\

\bottomrule
\end{longtable}

\end{document}